\documentclass{aa}
\usepackage{psfig}
\usepackage{lscape}
\usepackage{amsmath}
\usepackage{amssymb}
\usepackage[figuresleft]{rotating}

\newcommand{\MC}{\multicolumn}
\newcommand{\kms}{km~s$^\mathrm{-1}$}
\newcommand{\sunn}{$_{\odot}$}
\newcounter{qub}
\begin{document}

\title{
Starburst in HS~0822+3542 induced by the very blue LSB dwarf SAO~0822+3545
}

\author{S.A. Pustilnik\inst{1,4} \and
A.Y. Kniazev\inst{1,2,4} \and
A.G. Pramskij\inst{1,4}  \and
A.V. Ugryumov\inst{1,4} \and
J. Masegosa\inst{3}}

\offprints{S. Pustilnik, \email{sap@sao.ru}}

\institute{
Special Astrophysical Observatory RAS, Nizhnij Arkhyz,
Karachai-Circassia,  369167 Russia
\and Max Planck Institut f\"{u}r Astronomie, K\"{o}nigstuhl 17, D-69117,
Heidelberg, Germany
\and Instituto de Astrofisica de Andaluc\'ia, Granada, Spain
\and Isaac Newton Institute of Chile, SAO Branch
}

\date{Received \hskip 1cm February 26, 2002; accepted \hskip 1cm July 30, 2003}

 \abstract{
One of the most metal-deficient blue compact galaxies (BCGs) HS~0822+3542
($Z=1/34~Z$\sunn), is also one of the nearest such objects
($D \sim$11~Mpc). It is in addition well isolated from known bright
galaxies.
A trigger mechanism for its current star-formation (SF) burst has thus
remained unclear.
We report the discovery of a very blue
 ($(B-V)^0_{\rm tot}=$~0.08 and $(V-R)^0_{\rm tot}=$~0.14)
low surface brightness  (LSB) ($\mu_{B}^{0} \gtrsim$ 23\fm4~arcsec$^{-2}$)
dwarf irregular (dIrr) galaxy, which we have named SAO 0822+3545.
Its small relative velocity and projected distance of only $\sim$11
kpc from the BCG imply their physical association.  For this
LSB dIrr galaxy, we present spectroscopic results, total $B,V,R$
magnitudes, and the effective radii and surface brightness (SB), and we
describe its morphological properties.
We compare the very blue colours of this dwarf with  PEGASE.2 models of
the colour
evolution of a $Z=1/20~Z$\sunn\ stellar population, and combine this analysis
with the data on the LSBD $EW$(H$\alpha$) values.
The models best describing all available observational data depend on
the relative fraction of massive stars in the IMF used.
For a Salpeter IMF with $M_{\rm up}$ = 120 $M$\sunn, the best model
includes a ``young'' single stellar population (SSP) with an age of
$\sim$10~Myr and an ``old'' SSP with the age of $\sim$0.2--10~Gyr. The mass
ratio
of the old to young components should be in the range of 10 to 30.
If the age of the old component is more than $\sim$1 Gyr, an additional
coeval component of ``intermediate'' age ($\sim$100~Myr) with
a mass comparable to that of the ``young'' population, although  not
required, provided a good fit to the current data.
For the two options of a model IMF biased toward the low-mass end,
the best
match of the observed $BVR$ and EW(H$\alpha$) is for continuous
star-formation rate (SFR)
single-component models,  with SF durations in the range of $\sim$0.1 to
$\sim$1 Gyr. However, only a longer time-scale SF gives the stellar mass,
compatible with the LSB galaxy mass estimates. Nevertheless, such a scenario
would be inconsistent with the recent encounter of these two dwarfs.
The role of interaction between the LSBD and BCG HS~0822+3542 in triggering
their major SF episodes  during the last $\sim$100--200 Myr is emphasized and
discussed.  For the BCG, based on the results of new spectroscopy with the
Russian 6\,m telescope, we estimate the physical parameters of its SF
region and present the first evidence of an ionized gas supershell.
This pair of dwarfs lies deep within the nearby  Lynx-Cancer
void, with the nearest bright ($L > L_{*}$) galaxies at
distances $>$ 3 Mpc. This is probably one of the main factors
responsible for the unevolved state of HS 0822+3542.
   }

\authorrunning{S.A. Pustilnik et al.}

\titlerunning{Starburst in HS~0822+3542 induced by  SAO~0822+3545}

\maketitle

\section{Introduction}

A few known  blue compact galaxies (BCGs)
with extremely low metallicities (1/50 to
1/20 $Z$\sunn) are considered to be the best candidates for truly young local
low-mass galaxies, in which we are witnessing the first star formation
episode, with the oldest stars formed less than $\sim$100--200 Myr ago.
The best known
examples are SBS~0335$-$052 (Izotov et al. \cite{Izotov97}; Papaderos et al.
\cite{Papa98}; Pustilnik et al. \cite{PBTLI}),
and I~Zw~18 (Searle \& Sargent \cite{SS72}; Izotov \& Thuan \cite{IT99};
Papaderos et al. \cite{Papa02}).\footnote{Note that debates on the possible
youth of these BCGs still continue in the literature (see, e.g., the
most recent \"Ostlin
\cite{O_00}; \"Ostlin \& Kunth \cite{OK01}; Kunth \& \"Ostlin \cite{KO_01}).}
A question then arises: why did their progenitors -- protogalactic
\ion{H}{i} clouds
not experience gas collapse earlier, during the time after their
neutral gas settled down into the gravitational wells of their dark matter
halos?
Recent observational data indicate that some of them certainly interact with
their nearest neighbours (e.g., Dw 1225+0152, Chengalur et al.
\cite{Chengalur95}; SBS~0335$-$052~E, Pustilnik et al.~\cite{PBTLI}; and
probably IZw18 with the tiny object IZw18C, Dufour et al. \cite{Dufour96}, van
Zee et al. \cite{Zee98}, Izotov et al. \cite{Izotov01}). This, presumably,
is a key moment of their history. Moreover, it is probably
not by chance that in both the cases of Dw 1225+0152 and SBS 0335$-$052~E
the nearest neighbour is very gas-rich, and either is an \ion{H}{i} cloud
without any hint of past star formation (as in the case of Dw 1225+0152,
Salzer et al. \cite{Salzer91}), or is also an extremely metal-deficient,
probably truly young, blue compact galaxy (SBS 0335$-$052~W,
Pustilnik et al. \cite{Pustilnik97};  Lipovetsky et al. \cite{Lipovetsky99}).

Low surface brightness galaxies (LSBGs) comprise a large fraction of the
general field galaxy population and outnumber by a factor of several times the
high surface
brightness (HSB) population (e.g., McGaugh \cite{McGaugh96}, Dalcanton et al.
\cite{Dalcanton97}, O'Neil \& Bothun \cite{O'Neil00}).
Therefore, these LSBGs can be an important factor for interaction-induced
star-formation activity in gas-rich galaxies in general, mostly through
distant/weak tidal encounters (see, e.g., Taylor  et al. \cite{Taylor95};
O'Neil et al. \cite{O'Neil98}; Pustilnik et al. \cite{PKLU}). In particular,
through interaction they can trigger the first starbursts in the hypothetical
population of local protogalactic \ion{H}{i} clouds.
To check this hypothesis the authors are conducting a systematic study of the
local environment of the most metal-deficient BCGs. We present here
new evidence in support of this idea.
We report the discovery of a LSB dwarf irregular galaxy (named
SAO~0822+3545) at a projected distance of $\sim$11.4 kpc from
one of the most metal-deficient BCGs, HS~0822+3542 (Kniazev et al.
\cite{Kniazev00}).
We have used the SAO RAS 6\,m telescope spectrum in the H$\alpha$ region, as
well as $BVR$ photometry from the Nordic Optical Telescope (NOT)
to study its properties and estimate its tidal effect on
the BCG HS~0822+3542.
We also present new high signal-to-noise (S/N) 6\,m telescope spectra
for HS~0822+3542 which allow us to make more accurate measurements of some of
the physical parameters of its  star-forming (SF) region and
discover the first kinematic evidence of an
ionized gas supershell.
In Section~\ref{Obs} we describe the observations and their reduction. Results
of the data analysis are presented in Section~\ref{Results}. We discuss these
results in Section~\ref{Discussion},
and summarize our findings and draw conclusions in Section~\ref{Conc}.


   \begin{figure*}
   \centering
   \includegraphics[angle=-90,width=15cm]{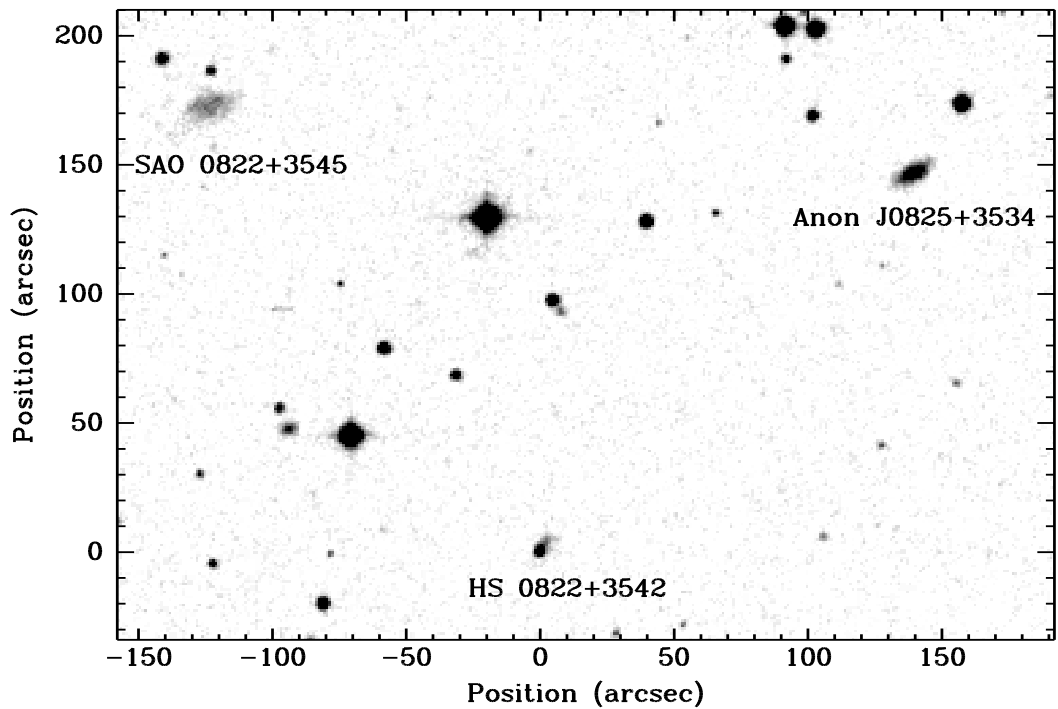}
      \caption{The Digitized Sky Survey (DSS-II) blue image of the field
               in the vicinity of BCG HS~0822+3542.
               North is up, East is to the left.
               At the adopted distance of HS~0822+3542 (11 Mpc) the scale
	       is $\sim$54 pc = 1\arcsec.
               The two marked galaxies -- LSBD SAO~0822+3545 and Anon
               J0825+3534 have been checked as possible companions of
	       the BCG. On our data the LSBD has $V_\mathrm{hel}
               \sim$700~\kms, close to that of HS~0822+3542.
               Its projected distance to the BCG is 213\arcsec\ or
	       11.4 kpc. The galaxy Anon J0825+3534, very
	       close to the position of 2MASXi J0825440+353459 (NED),
	       is a distant background object with
	       $V_\mathrm{hel}=$15590~\kms.
	       }
         \label{FigDSS}
   \end{figure*}

\section{Observations and reduction}
\label{Obs}


\begin{table*}
\begin{center}
\caption{Journal of the 6\,m telescope spectroscopic observations}
\label{Tab1}
\begin{tabular}{lrrrccccc} \\ \hline \hline
\MC{1}{c}{ Object }     &
\MC{1}{c}{ Date }       &
\MC{1}{c}{ Exposure }   &
\MC{1}{c}{ Wavelength } &
\MC{1}{c}{ Dispersion } &
\MC{1}{c}{ Seeing }     &
\MC{1}{c}{ Airmass }     &
\MC{1}{c}{ PA }          \\

\MC{1}{c}{ }       &
\MC{1}{c}{ }       &
\MC{1}{c}{ time [s] }    &
\MC{1}{c}{ Range [\AA] } &
\MC{1}{c}{ [\AA/pixel] } &
\MC{1}{c}{ [arcsec] }    &
\MC{1}{c}{          }    &
\MC{1}{c}{ [degree] }     \\

\MC{1}{c}{ (1) } &
\MC{1}{c}{ (2) } &
\MC{1}{c}{ (3) } &
\MC{1}{c}{ (4) } &
\MC{1}{c}{ (5) } &
\MC{1}{c}{ (6) } &
\MC{1}{c}{ (7) } &
\MC{1}{c}{ (8) } & \\
\hline
\\[-0.3cm]
HS~0822+3542 & 12.02.1999  & 2x1800 & $ 3700-8000$  & 4.6 &  1.8 & 1.10 & 142 \\
HS~0822+3542 & 13.02.1999  & 2x1800 & $ 3700-8000$  & 4.6 &  1.4 & 1.31 & 142 \\
HS~0822+3542 & 01.11.2000  & 2x1800 & $ 6200-7400$  & 1.2 &  1.7 & 1.02 & 175 \\
SAO~0822+3545 & 17.01.2001  & 1800 & $ 5700-8100$  & 2.4 &  2.0 & 1.03 &  134 \\
Anon J0825+3534 & 02.02.2000  &  300 & $ 5700-8100$  & 2.4 & 2.4 & 1.09 &  121 \\
\hline \hline \\[-0.2cm]
\end{tabular}
\end{center}
\end{table*}

\subsection{Nordic Optical Telescope photometry}
\label{NOT}

There are no cataloged galaxies around HS~0822+3542 in either NED or LEDA
databases with $V_\mathrm{hel} <$ 1200~\kms\ and projected distances less
than 5.8\degr. To  search for fainter/non-cataloged galaxies in the close
vicinity of this BCG we used $B,V,R$ CCD images obtained with the 2.5\,m
NOT on May 28, 1998. These are the same frames which were used to derive
$B,V,R$ magnitudes for HS~0822+3542. They were acquired
with the ALFOSC spectrograph equipped with the LORAL (W11-3AC) CCD,  which
has a direct imaging mode. Exposure times were 900~s for the $B$-band
image, and 600~s for both $V$ and $R$ images. For further details of these
observations and data reduction we refer to the paper by Kniazev et
al. (\cite{Kniazev00}).

\subsection{Looking for possible companions}

The photometric data were reduced with MIDAS\footnote{MIDAS is an
acronym for the European Southern Observatory package -- Munich Image Data
Analysis System.} Command Language programs according to the method
described in Kniazev (\cite{Kniazev97}). The MIDAS {\tt INVENTORY} package
was used to classify all objects. Isophotal and total $BVR$ magnitudes
were computed using the transformation coefficients of Kniazev et al.
(\cite{Kniazev00}).

Since the interaction between HS~0822+3542 and a possible
neighbouring galaxy would probably result in some enhanced star formation in
the latter, we first of all searched for candidate blue galaxies.
The second criterion applied  was that the brightness of any candidates
should be comparable to that of HS~0822+3542, since significantly fainter
neighbours could tidally affect the BCG only from very small distances (see,
e.g., estimates in Pustilnik et al. \cite{PKLU}).

Only two blue galaxies were found in the examined field. The first one
is HS~0822+3542 itself. The second is a LSB
irregular galaxy (SAO~0822+3545) at 3.5\arcmin\ to the north-east from
HS~0822+3542 (its coordinates are given in Table \ref{t:Param}).
The Digitized Sky Survey (DSS-II) blue image of this field with the 2 objects
of interest and  Anon J0825+3534 is presented in Fig.~\ref{FigDSS}.
All other galaxies in the NOT field were either significantly fainter or
redder.

\subsection{Analysis of photometric data}
\label{Photometry}

We performed the reduction of photometric data for
SAO 0822+3545 using the
IRAF\footnote{IRAF: the Image Reduction and Analysis
Facility is
distributed by the National Optical Astronomy Observatories, which is
operated by the Association of Universities for Research in Astronomy,
In. (AURA) under cooperative agreement with the National Science
Foundation (NSF).}
package {\tt ELLIPSE}.
The growth curve (GC) of the galaxy was constructed
by summing up the pixel values from the center outwards
in the circles of successive radius.

The total magnitudes ($B_\mathrm{tot}$, $V_\mathrm{tot}$, $R_\mathrm{tot}$)
were estimated by asymptotic extrapolation of the respective radial GCs.
Model-independent parameters were derived from each growth curve.
The effective radii ($r_\mathrm{eff}$) were read on each GC at one
half the asymptotic intensity, and effective SBs ($SB_\mathrm{eff}$)
were determined as the mean brightness within a circle with the effective
radius. These values are summarized in Table~\ref{tab:struct_par}.


   \begin{figure}
   \centering
   \includegraphics[width=8.5cm]{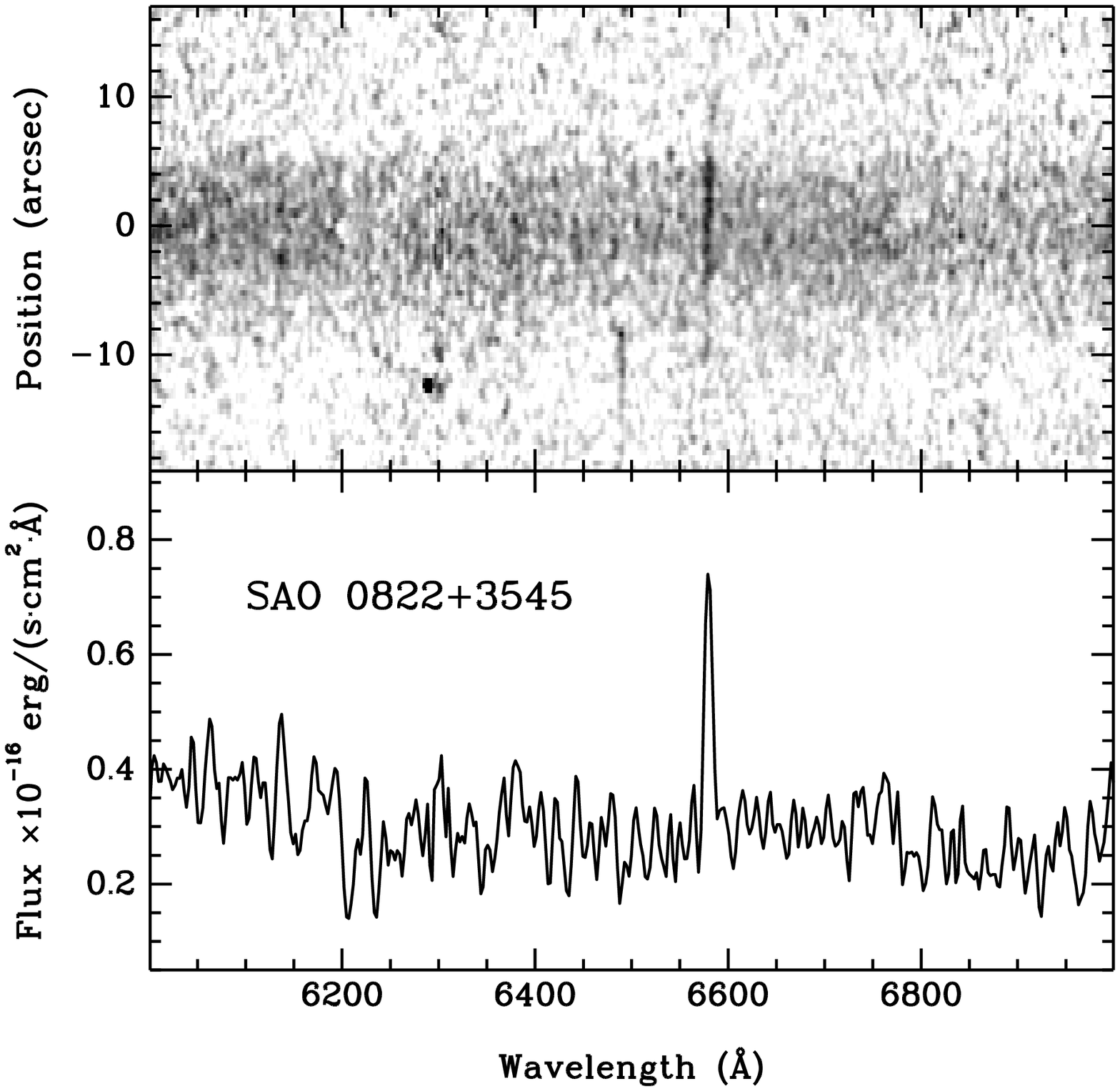}
    \caption{{\it Top panel:} 2D spectrum of SAO~0822+3545 with the H$\alpha$
       line at $\lambda$6579~\AA.
       Positive $Y$ corresponds to the NW direction for the slit position
       shown in Fig.~\ref{LSB_B_image}.
       H$\alpha$-emission  is clearly seen on top of the continuum in the
       region of $\pm \sim$5\arcsec\ from the central position.
       Furthermore, some very faint H$\alpha$-emission (with S/N ratio
       of $\sim$1) can be traced out to
       $\pm \sim$10\arcsec, in the regions without detectable continuum.
       {\it Bottom panel:} 1D spectrum of the central region ($\pm$3\arcsec),
       summed along the slit, $EW$(H$\alpha$) = 17~\AA.
        }
         \label{fig:LSBHa}
   \end{figure}

\subsection{Long-slit spectroscopy}

Spectroscopic data were obtained  with the 6\,m telescope of the Special
Astrophysical Observatory of Russian Academy of Sciences (SAO RAS). The
long-slit spectrograph (LSS) (Afanasiev et al. \cite{Afanasiev95})
was used with a Photometrics 1024$\times$1024 pixel CCD detector  with a
24~$\mu$m pixel size. The wavelength ranges of the spectra obtained
for different hardware configurations are given in Table~\ref{Tab1}.
Reference spectra of an Ar--Ne--He lamp were recorded before or after
each observation to provide  wavelength calibration. All observations were
conducted mainly with the software package {\tt NICE} in MIDAS,
described by Kniazev \& Shergin (\cite{Kniazev95}).

H$\alpha$ emission from SAO~0822+3545 was detected in observations on
January 17, 2001, with a 651 groove~mm$^{-1}$ grating and a slit
width of 2\arcsec\ (along the major axis), resulting in
2.4~\AA~pixel$^{-1}$ and   a full width at half maximum (FWHM)
resolution of $\sim$7~\AA.

Two high S/N spectra of HS~0822+3542 with the 325 groove~mm$^{-1}$ grating,
giving a sampling of 4.6~\AA~pixel$^{-1}$ and an effective resolution of
$\sim$14~\AA\ (FWHM), were obtained during 2 photometric nights on February 12
and 13, 1999. Seeing on these nights was  $\sim$1.8\arcsec\ and
$\sim$1.4\arcsec, respectively.
To keep the long slit (1.2\arcsec$\times$120\arcsec) at the same position
and $PA$ (position angle), we used the method of differential pointing,
similar to that
described by Kniazev et al. (\cite{UM}). Spectrophotometric standard stars
from Bohlin (\cite{Bohlin96}) were observed for flux calibration.

The radial velocity  distribution of ionized gas along the slit, as seen
in the H$\alpha$-line (Position--Velocity or \mbox{P--V} diagram) was derived
for HS~0822+3542 from the LSS observations on November 1, 1999, with the
1302 groove~mm$^{-1}$ grating, giving a sampling of  1.2~\AA~pixel$^{-1}$
and an effective
resolution (with the 2\arcsec\ slit width) of 3.5~\AA. Seeing during these
observations was 1.7\arcsec.

For the reduction of 2D spectra we obtained biases,  flat fields and
illumination correction images. The primary reduction consisted of
standard steps and was done using the IRAF package {\tt CCDRED}.
The IRAF package {\tt LONGSLIT} was used to perform wavelength
calibration, background subtraction, extinction correction and flux
calibration. Straightening of the 2D spectra was performed using {\tt APALL},
another IRAF package. After that, aperture extractions, continuum
determination,
flux and equivalent width measurements of spectral lines  were performed in
MIDAS (for details, see Kniazev et al.~\cite{Kniazev00}).
The reduced 2D spectrum and extracted averaged 1D spectrum of
SAO 0822+3545 are shown in Fig.~\ref{fig:LSBHa}.

Fluxes and equivalent widths of blended lines were measured using Gaussian
decomposition fitting. An average sensitivity curve was produced for each
night
with r.m.s. deviations of~~$\lesssim$5\% in the whole spectral range.
The sensitivity curve and line intensity errors have been propagated in
calculating elemental abundances. To construct the P--V diagram, the
methodology described in Zasov et al. (\cite{Zasov00}) was used,
which allowed us to measure points with sufficiently bright
H$\alpha$ emission with r.m.s. errors on the level of 2--4~\kms.


   \begin{figure*}
   \centering
   \includegraphics[width=8.5cm]{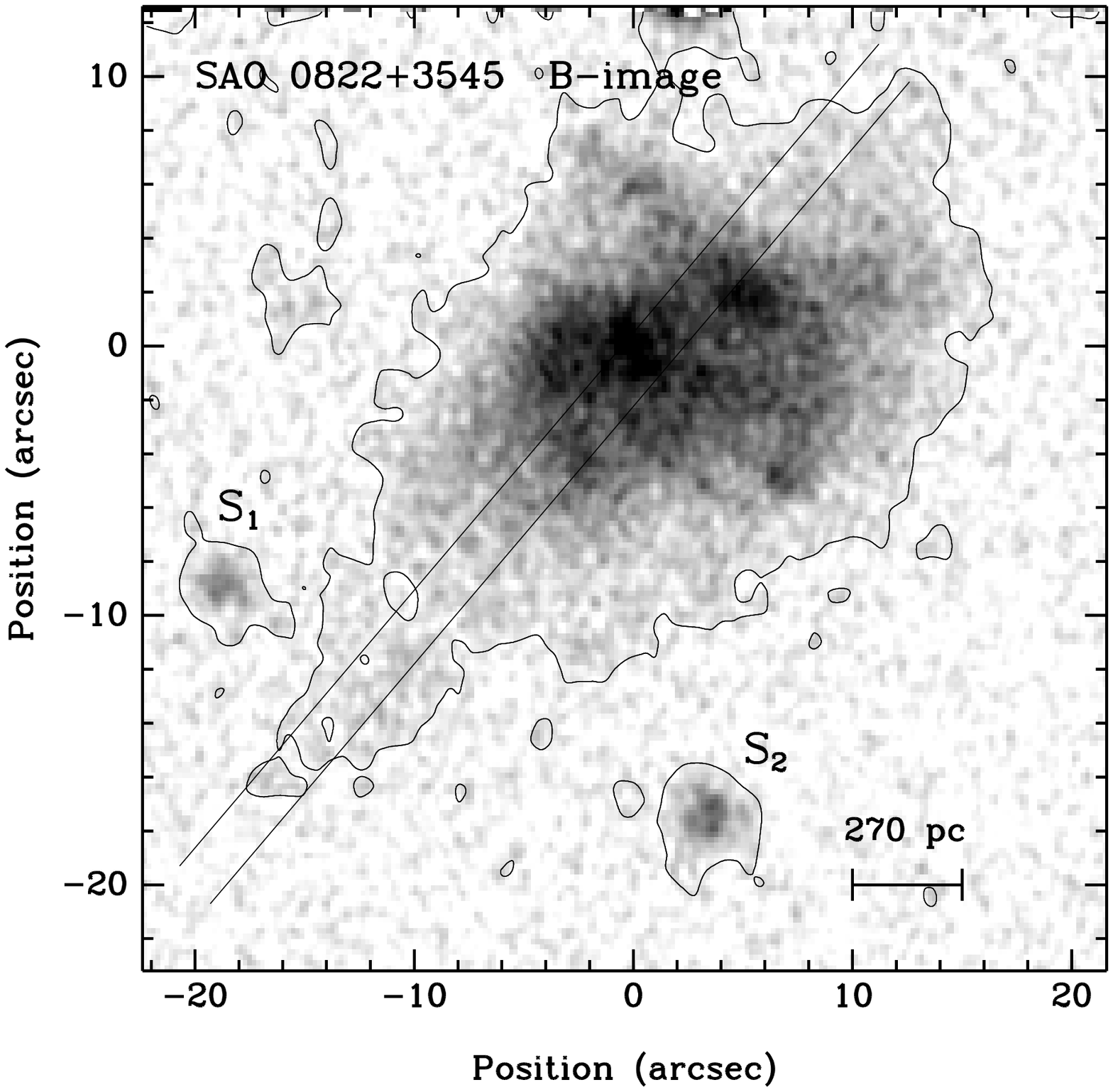}
   \includegraphics[width=8.5cm]{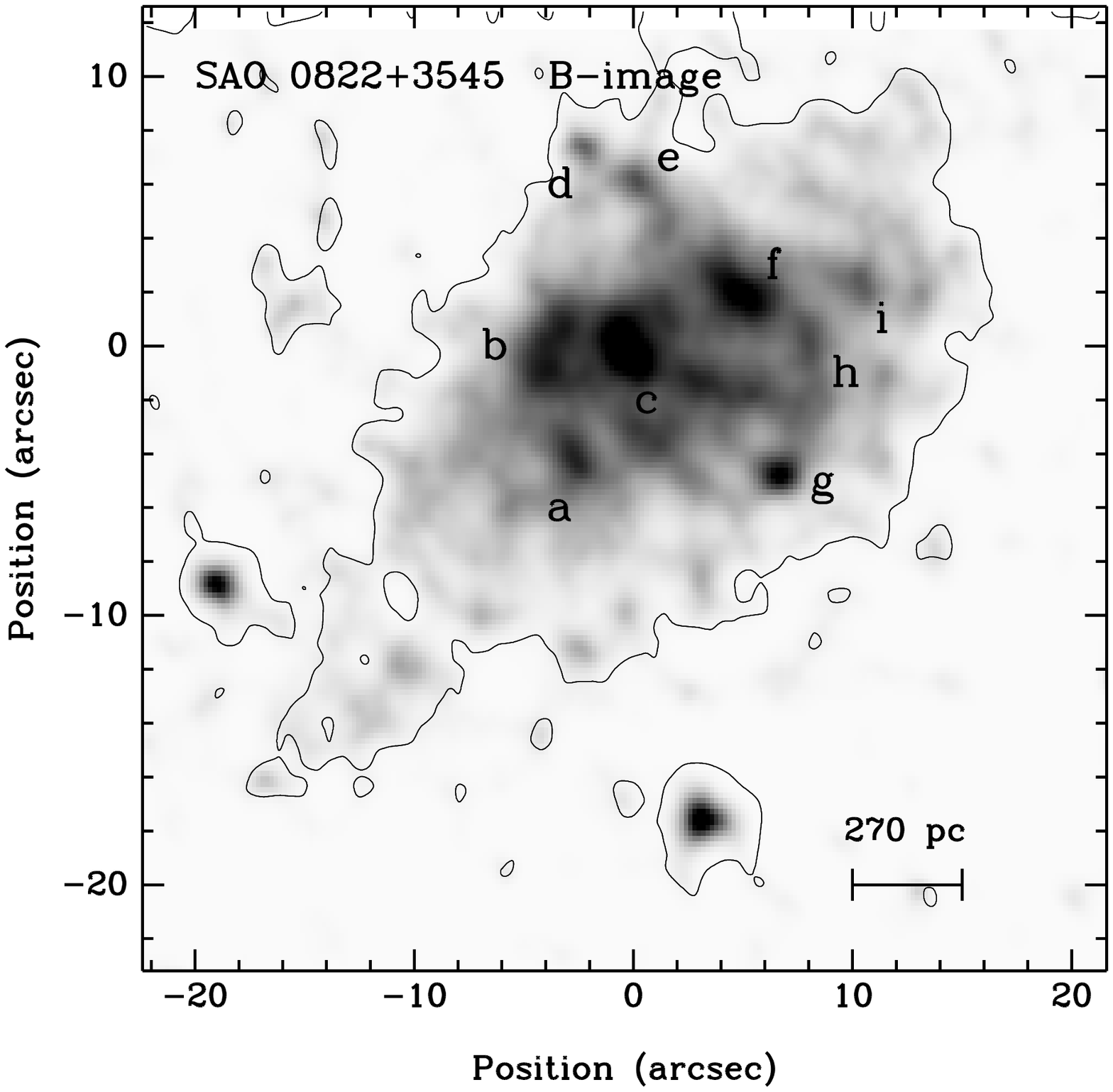}
    \caption{\underline{\it Left panel}: $B$-band image of SAO~0822+3545
       from the NOT. N is up, E is to the left.
       The 2\arcsec-wide long slit of the 6\,m telescope spectrum is
       superimposed.
       Highly irregular structure is seen in the inner $\sim$15\arcsec.
       Two knots near the center and two arcs to the
       N and W are indicative of some activity in this galaxy. The isophotal
       contour corresponds to a $B$-band SB of $\sim$25\fm7~arcsec$^{-2}$, or
       2$\sigma$ of background (for 2D Gaussian smoothing with FWHM=1\arcsec).
       Two faint objects, probable background galaxies in the vicinity of
       SAO 0822+3545, are marked as S$_{\rm 1}$ and S$_{\rm 2}$.
       \underline{\it Right panel}: The same $B$-band image of
       SAO 0822+3545, deconvolved after 15 iterations using the  FFT
       based Lucy algorithm  (Adorf et al.~\cite{Adorf92}).
       The angular resolution has improved from 1\farcs25 to
       $\sim$0\farcs85\ (FWHM). Several filaments and many knots are easily
       visible within the brighter part of the galaxy. Knots are marked by
       letters from ``a'' to ``i'', respectively.
    }
         \label{LSB_B_image}
   \end{figure*}

\section{Results}
\label{Results}

\subsection{Spectrum of SAO~0822+3545}
\label{LSB_spectrum}

The 2D spectrum of SAO~0822+3545  in Fig.~\ref{fig:LSBHa} shows H$\alpha$
emission spanning $\sim$10\arcsec, comparable to the apparent extent
of the galaxy continuum. Some fainter H$\alpha$ features can be detected in
the peripheral regions, where no continuum is seen in our spectrum.  The
equivalent width of the H$\alpha$ line, measured on the 1D spectrum averaged
over the central region with the a spatial extent of 6\arcsec, is
$EW$(H$\alpha$)=17$\pm$2~\AA. The integrated flux in this line is
4.65($\pm$0.55)$\times$10$^{-16}$ erg~s$^{-1}$~cm$^{-2}$.
Small variations of $EW$(H$\alpha$) along the slit near the center of
the galaxy on the level of $\sim15-20$~\AA\ are within the observational
uncertainties. Closer to the edges the $EW$(H$\alpha$) can be larger, but
there the uncertainty reaches 70--100\%, so we do not discuss these regions
further.
The measured velocity
$V_{\rm hel} =$700$\pm$50~\kms\ is very  close to that of HS~0822+3542.

\subsection{Photometry and morphology of SAO~0822+3545}
\label{Morphology}

We calculated total magnitudes $B_{\rm tot}$ = 17.56$\pm$0.05,
$V_{\rm tot}$ = 17.43$\pm$0.03 and
$R_{\rm tot}$ = 17.26$\pm$0.05 for SAO~0822+3545 using the growth
curve method as described in Sec.~\ref{Photometry}.
The respective integrated colours are $(B-V)_{\rm tot}$ = 0.13
and $(V-R)_{\rm tot}$ = 0.17.
Relatively large  errors originate from the zero-point uncertainties of
the transformation equations (Kniazev et al.~\cite{Kniazev00}).
Accounting for a foreground extinction of
$E(B-V)$ = 0.047 in our Galaxy (Schlegel et al. \cite{Schlegel98}),
and applying the  extinction curve from Whitford (\cite{Whitford58}),
these colours become somewhat bluer:
$(B-V)_{\rm tot}^{0}$ = 0.08$\pm$0.06,
$(V-R)_{\rm tot}^{0}$ = 0.14$\pm$0.06.

The 2D spectrum in Fig.~\ref{fig:LSBHa} and the irregular structure
of the central part of SAO 0822+3545 (see Fig.~\ref{LSB_B_image}) indicate
that enhanced  star formation  took place at least in the inner part of that
galaxy. The possibility of better distinguishing individual regions of higher
brightness
(possible young superclusters and aged \ion{H}{ii} regions) in the central
part of the LSBD was the motivation for using the MIDAS package {\tt IMRES}
for deconvolution of the NOT images.
This package employs an image restoration scheme devised by Richardson and
Lucy and
described in Adorf et al.~(\cite{Adorf92}),  Hook \& Lucy~(\cite{Hook_Lucy92};
and references therein).
The number of iterations was determined by comparisons between the input
image and the output deconvolved images, after convolution with the
point-spread function (PSF).

The results of deconvolving the $B$-band image after 15 iterations are plotted
in the right panel of Fig.~\ref{LSB_B_image}. A number of filaments and
relatively bright knots in the inner part of this LSB/dIrr galaxy are well
resolved. Practically all of them are seen on the original image
(left panel of Fig.~\ref{LSB_B_image}), although deconvolution makes
the structure more visible.
All these structures are easily visible in a deconvolved $V$-band
image, but are less distinct in $R$-band.

There is a faint tail in the SE direction in the outer part of the LSBD.
This tail is visible on both blue DSS-II and $B$-band NOT images. The
surface brightness of this tail corresponds to $\mu_B =$ 25\fm7~arcsec$^{-2}$,
or 2$\sigma$ of the noise level for the $B$-band image.  The tail direction
coincides with that of the LSBD major axis. The major axis $PA$ varies
from --75\degr\ for the inner part of the galaxy up to --60\degr\ for the
outer part. The axial ratio is essentially constant over the body:
$b/a \sim$0.55.

\begin{figure}
   \includegraphics[width=8.5cm]{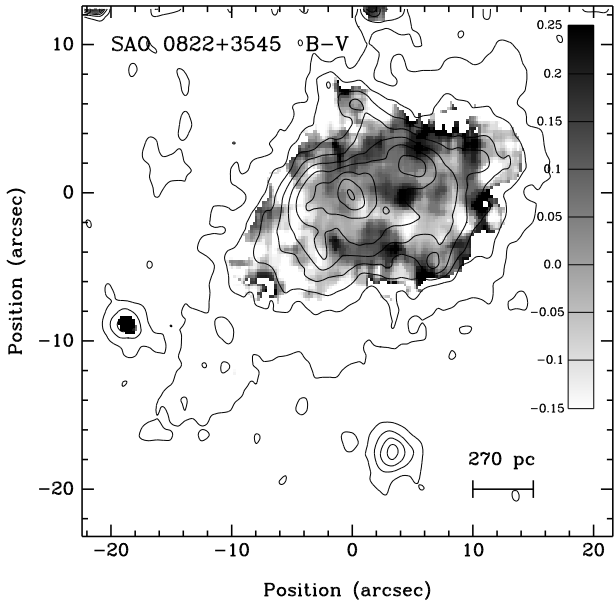}
    \caption{
     Grey-scale map of $(B-V)$-colour for SAO 0822+3545 smoothed with
     2D median with the window size of 1\farcs6$\times$1\farcs6.
     The extinction corrected ($E(B-V)$=0.047) colour distribution is
     shown only for regions with S/N ratio larger than 4.
     The colour scale is shown in the right side column. $B$-band surface
     brightness (SB) isolines are superimposed, with the lowest SB level
     of 25\fm7~arcsec$^{-2}$. The other levels are: 24.5, 24.0, 23.6, 23.3,
     23.1, 22.9, 22.8 and 22.6 mag~arcsec$^{-2}$.
        }
         \label{LSB_B-V_image}
   \end{figure}

Evidence for recent star-formation activity over the internal part of this
dIrr galaxy is also seen, e.g.,  in the $(B-V)_{0}$ colour map
(Fig.~\ref{LSB_B-V_image}).
This  map shows a rather uniform distribution (mainly in the range
from --0.05 to +0.10) over a large part of the main body.
Its maximum  (+0.2 to +0.25) and  minimum (--0.15 to --0.10)
values are measured in the positions close to the  edges,
where the S/N is small, and these large colour variations are apparently
spurious.

SAO 0822+3545 is a genuine LSB galaxy with traces of recent SF near its
center. Since the central bright knots are seen with the lowest contrast
in $R$-band, we made a rough estimate of the central SB of an underlying
``disk'' in $R$. From the data in Table~\ref{tab:struct_par},
its effective surface brightness corrected for extinction ($A_{R}$=0.12) is
$\mu_{\rm eff,0}^{R}$=23\fm63~arcsec$^{-2}$. For a purely exponential disk
this corresponds to a central brightness of
$\mu_{\rm 0}^{R}$=22\fm50~arcsec$^{-2}$.
Since the galaxy is significantly inclined to the line-of-sight, we need to
make a corresponding correction to $\mu_{\rm 0}^{R}$.
For an observed axial ratio $p=b/a=0.55$, assuming an intrinsic axial ratio of
$q=$0.2, and using the well-known formula
$\cos^{2}(i)=(p^{2}-q^{2})/(1-q^{2})$, we calculate that $i=58.5$\degr.
The inclination correction for surface brightness is then
\mbox{$-2.5\cdot\log$($\cos~i$)=0\fm7}, giving a corrected central brightness
of $\mu_{\rm 0,c}^{R}$=23\fm20 arcsec$^{-2}$.
Even if the underlying ``disk'' is as blue, as the integrated light of this
galaxy (that is, $(B-R)_{\rm 0} \sim$0.2), its central brightness in $B$-band
$\mu_{\rm 0,c}^{B}$=23\fm40 arcsec$^{-2}$ is well within the LSB galaxy
regime.
For an exponential law, $r_{\rm eff}$=7.94\arcsec\ in $R$ corresponds to
the scalelength $\alpha_{\rm R}$=4.73\arcsec.

\begin{table}[hbtp]
\caption{Photometric parameters of SAO 0822+3545}
\label{tab:struct_par}
\begin{tabular}{cccc} \hline \hline
Band&  Total          &$r_\mathrm{eff}$&$SB_\mathrm{eff}$ \\
    &   mag           &    arcsec      &mag arcsec$^{-2}$ \\
    &   (1)           &     (2)        &     (3)          \\ \hline
$B$ & 17.56$\pm$0.05  & 6.46~          &  23.60          \\
$V$ & 17.43$\pm$0.03  & 6.43~          &  23.46          \\
$R$ & 17.26$\pm$0.05  & 7.94~          &  23.75          \\
\hline\hline
\end{tabular}
\end{table}


\subsection{Morphology and ionized gas motions of HS~0822+3542}
\label{curve}

\begin{figure*}
   \includegraphics[width=8.5cm]{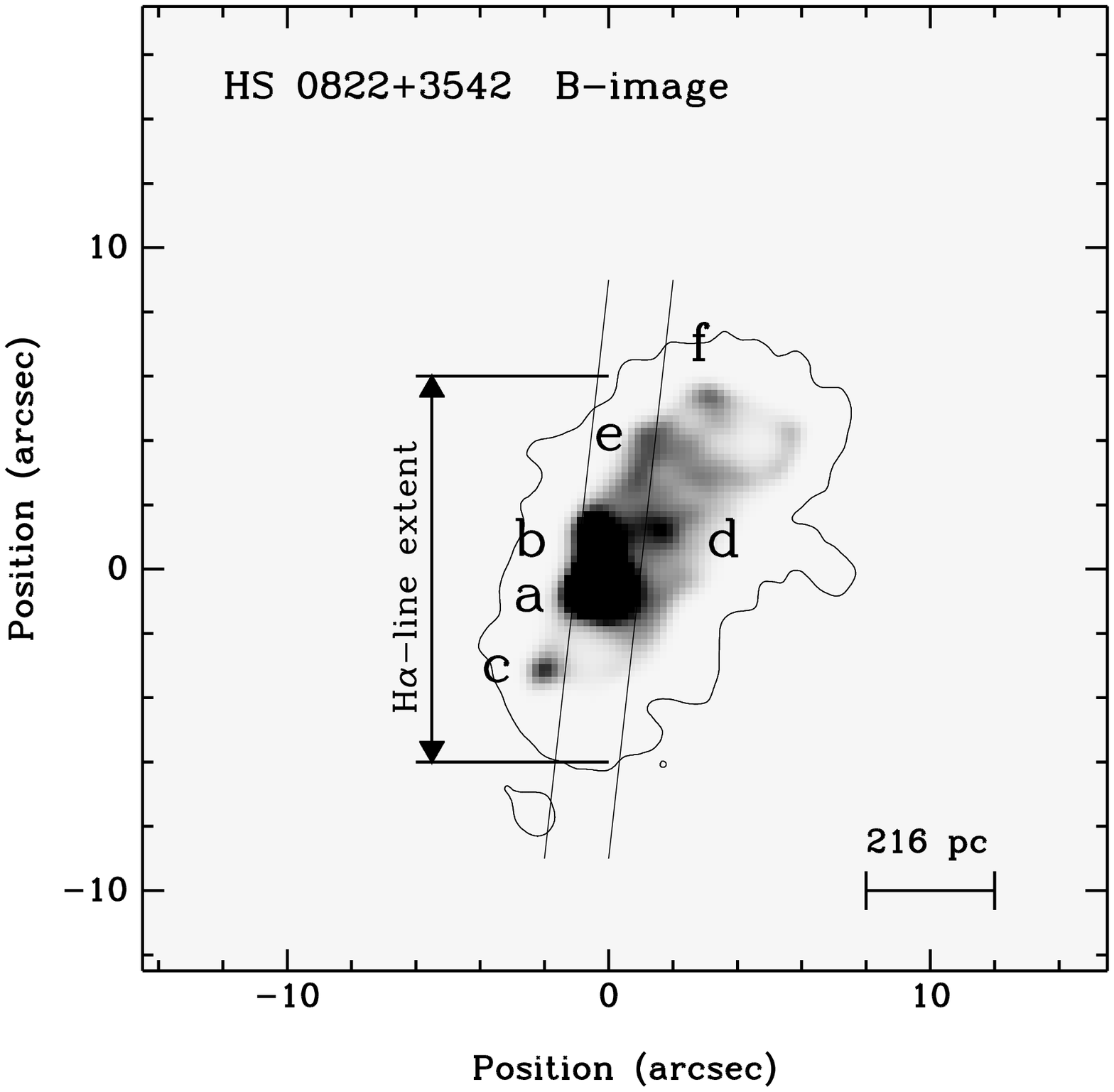}
   \includegraphics[width=8.5cm]{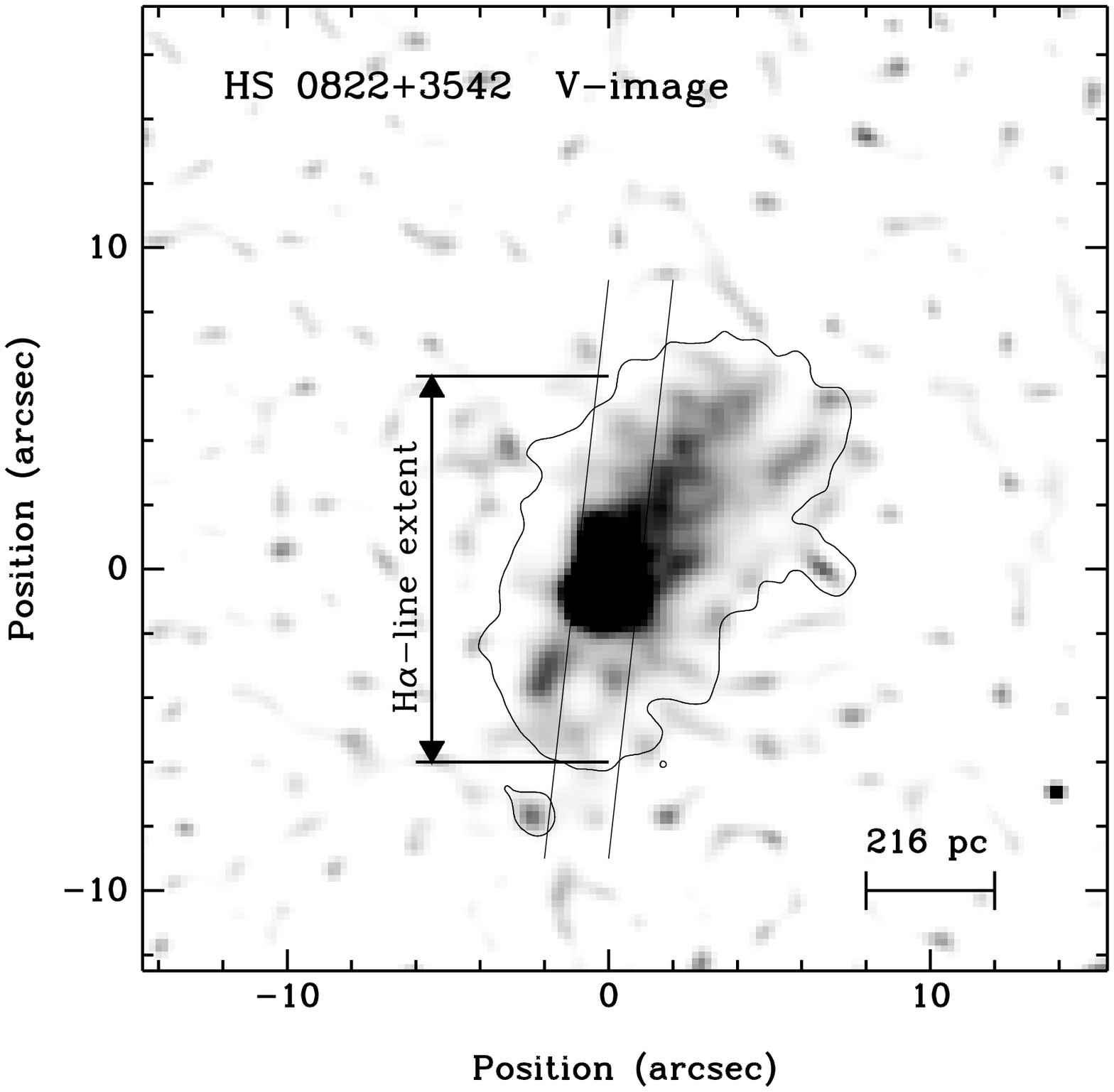}
    \caption{\underline{\it Left panel:}  The NOT $B$-band image of
       HS~0822+3542, deconvolved with
       40 iterations using FFT based Lucy algorithm. The angular resolution
       is improved from 1\farcs25 to $\sim$0\farcs6. N is up, E is to the
       left. The slit position for H$\alpha$ velocity curve is superimposed.
       The isophotal contour corresponds to $\mu_{\rm B} =$
       25\fm0~arcsec$^{-2}$ or 4$\sigma$ of background (for 2D Gaussian
       smoothing with FWHM=1\arcsec). The arrow indicates the extent of
       H$\alpha$ emission on the 6\,m telescope spectrum. All features marked
       as ``a'' to ``f'' are resolved in $B,V,R$ deconvolved images.
       Knot ``b'' is $\sim$8 time fainter than ``a'' (see section
       \ref{curve}). \underline{\it Right panel:} Same as before, but for
       NOT $V$-band image. The filamentary structure on NW edge is well seen
       on both images.
    }
         \label{HS0822_B_deconv}
\end{figure*}

We also applied the image restoration method to the NOT images of
HS~0822+3542. After 40 iterations the angular resolution of the $B$-band image
was improved by a factor of two
(FWHM=0\farcs6), and the restored BCG complex morphology is shown in
Fig.~\ref{HS0822_B_deconv} in grey scale, with a contour indicating the outer
isophote
$\mu_{\rm B}=$25\fm0~arcsec$^{-2}$ superimposed. With the letters ``a''
to ``f'' we have marked all bright features well resolved after
deconvolution in each of $B,V$, and $R$ bands. The brightest region consists
of two components,
``a'' and ``b'', separated by $\sim$1.3\arcsec\
($\sim$70 pc), roughly in the N--S direction.
The contrast in Fig.~\ref{HS0822_B_deconv} is adjusted to
show faint features. The real ratio of the intensities of knots ``a'' and
``b'', derived from this image, is $\sim$8. While deconvolution does not
preserve brightness proportions between individual features, we consider
this ratio to be indicative of the real value.

In addition to several bright knots, some filamentary structures are also
visible. Some traces of them are also seen in other filters, but
due to their lower surface brightness
they are not as easily visible as the knots.
Nevertheless, the prominent arc-like
structure at the NW edge of the main body  (stretching from $X$=+2,
$Y$=+4 to $X$=+6, $Y$=+4 in Fig.~\ref{HS0822_B_deconv}), is clearly
visible on both $B$ and $V$ images, but is more noisy in $R$-band, presumably
due to the lower S/N ratio.
This feature probably represents
an ionized gas shell with a diameter of $\sim$4\arcsec\
($\sim$200 pc),  caused by recent active SF in this region,
Four more knots are seen in the restored image. One (``d'') is near the
geometrical center of the main body. This, perhaps, might naturally be
expected because of
the gravitational well in the center of the galaxy.
The others knots provide further evidence
that the current SF episode in the BCG is spread across the galaxy.

\begin{figure}
     \includegraphics[width=8.0cm,bb= 40 165 585 720,clip=]{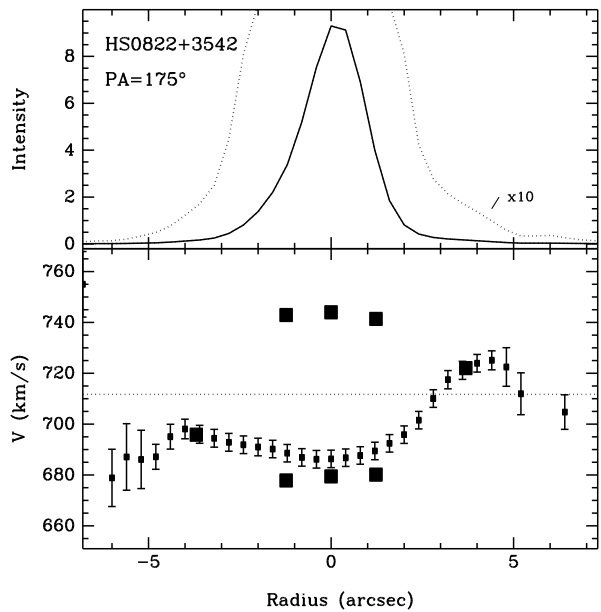}%
    \caption{\underline{\it Upper panel:} H$\alpha$ intensity in HS
    0822+3542 along the slit position shown in Fig.~\ref{HS0822_B_deconv}.
    \underline{\it Lower panel:} P--V diagram of ionized gas in HS~0822+3542
    based on the H$\alpha$ line from the same spectrum. Its form within a
    radius of $\pm$4\arcsec, and the characteristic diameter and
    velocity amplitude ($\sim$480 pc and $\sim$30~\kms), are consistent
    with the appearance of supershells observed in many star-bursting
    galaxies. Filled boxes show the velocities on both sides of the shell as
    derived from the Gaussian decomposition of the H$\alpha$ line at several
    positions along the slit. For details see section~\ref{curve}.
        }
         \label{fig:RotCurv}
   \end{figure}

The P--V diagram in Fig.~\ref{fig:RotCurv}
indicates a supershell with a size of $\sim$480 pc, comparable to the
extent of H$\alpha$-emission in the long-slit direction. The supershell
velocity amplitude, as seen from our data, is about 30~\kms.
Such supershells are well-resolved on high-resolution H$\alpha$ long-slit
spectra for many nearby
starbursting galaxies (e.g., Marlowe et al. \cite{Marlowe95}; Martin
\cite{Martin96,Martin97,Martin98}). The related supershells of neutral gas
are also seen in \ion{H}{i} maps of such galaxies (e.g., Walter \& Brinks
\cite{WB99}).
For the typical case of a starburst off-set in position relative to the
midplane of the gas disk,
the asymmetry of the gas density distribution in the $z$-direction results
in the shell's asymmetric appearance. This was shown, e.g., in numerical
simulations by Silich et al. (\cite{Silich96}), and by Walter \&
Brinks (\cite{WB99}) for the observed types of P--V diagrams.
That part of the shell propagating out of the plane should have a
significantly
lower Emission Measure due to decreasing  gas density, and usually appears
much fainter in comparison to the part of the shell moving towards the
midplane (e.g., Martin \cite{Martin96}).

Shells (or supershells) are produced by hot bubbles, caused by the
injection of the energy from numerous massive star winds and supernova (SN)
explosions into the interstellar medium (ISM) (e.g., Tenorio-Tagle \&
Bodenheimer \cite{TTB88}). The
asymmetric appearance of supershells in H$\alpha$ emission is easily seen
with the high-resolution data mentioned above. When the velocity resolution
is not sufficient to distinguish motions on both sides of the shell,
we measure the intensity-weighted velocity in the H$\alpha$-line at each slit
position. If the H$\alpha$ intensities in the segments on opposite sides
of the shell differ significantly, we will see mainly the side  approaching
the midplane of the gas disk.
However, since the observed velocity at each slit position is the weighted
mean of the emission from the opposite sides of the shell, a shell
velocity derived in this way represents a lower limit of the real value.
Depending on the relative strengths of H$\alpha$ emission on the shell sides
the expected correction could reach tens of percent.

If we accept the full amplitude of the radial velocity difference
between the two edges  on the P--V diagram
($\sim$35~\kms) as the result of rotation in the BCG,
then the apparent $V_{\rm rot}$ is 17.5~\kms. With an apparent axial ratio
$p=b/a$=0.5, and an assumed intrinsic ratio $q=0.2$, using
the same relation as for SAO 0822+3545 in section \ref{Morphology} we get
an inclination angle $i$=63.6\degr, and thus an inclination-corrected value
of $V_{\rm rot}$=19.5~\kms. The latter value is quite consistent with that
expected from the Tully-Fisher relation between galaxy $V_{\rm rot}$ and its
blue luminosity, namely, using the relations derived by Karachentsev et al.
(\cite{KMH99}) for dwarf galaxies in the Local Volume. For HS~0822+3542,
$M_{\rm B}$=--12.5 (or $L_{\rm B}$=1.56$\times$10$^{7}$~$L$\sunn)
and so the expected $V_{\rm rot}=$17~\kms, with a $\pm$1$\sigma$ confidence
range of 12 to 24~\kms.

In Fig.~\ref{fig:RotCurv} we use filled squares to show the velocities on
both sides of the shell, obtained by Gaussian decomposition of the H$\alpha$
profiles. The two-component structure is detectable only for sufficiently
large S/N ratios and maximal velocity separation. The flux from the receding
components is several times lower than that from the approaching ones. These
data confirm the existence of a large shell with a characteristic velocity
amplitude of $\sim$30~\kms\ and exclude the interpretation of this feature
in the P--V diagram as part of a rotation curve.
Indeed, the rotation velocity should monotonously increase from one edge
to another. The real P--V diagram certainly does not look like this.
Thus, if the ionized gas in the BCG is rotating, its rotation velocity
is sufficiently small to be practically hidden by the visible large shell.

\subsection{Chemical abundances in HS~0822+3542}
\label{abun}

\begin{figure*}[tbh]

   \begin{minipage}[t]{\linewidth}
   \centering
   \includegraphics[angle=-90,width=16cm]{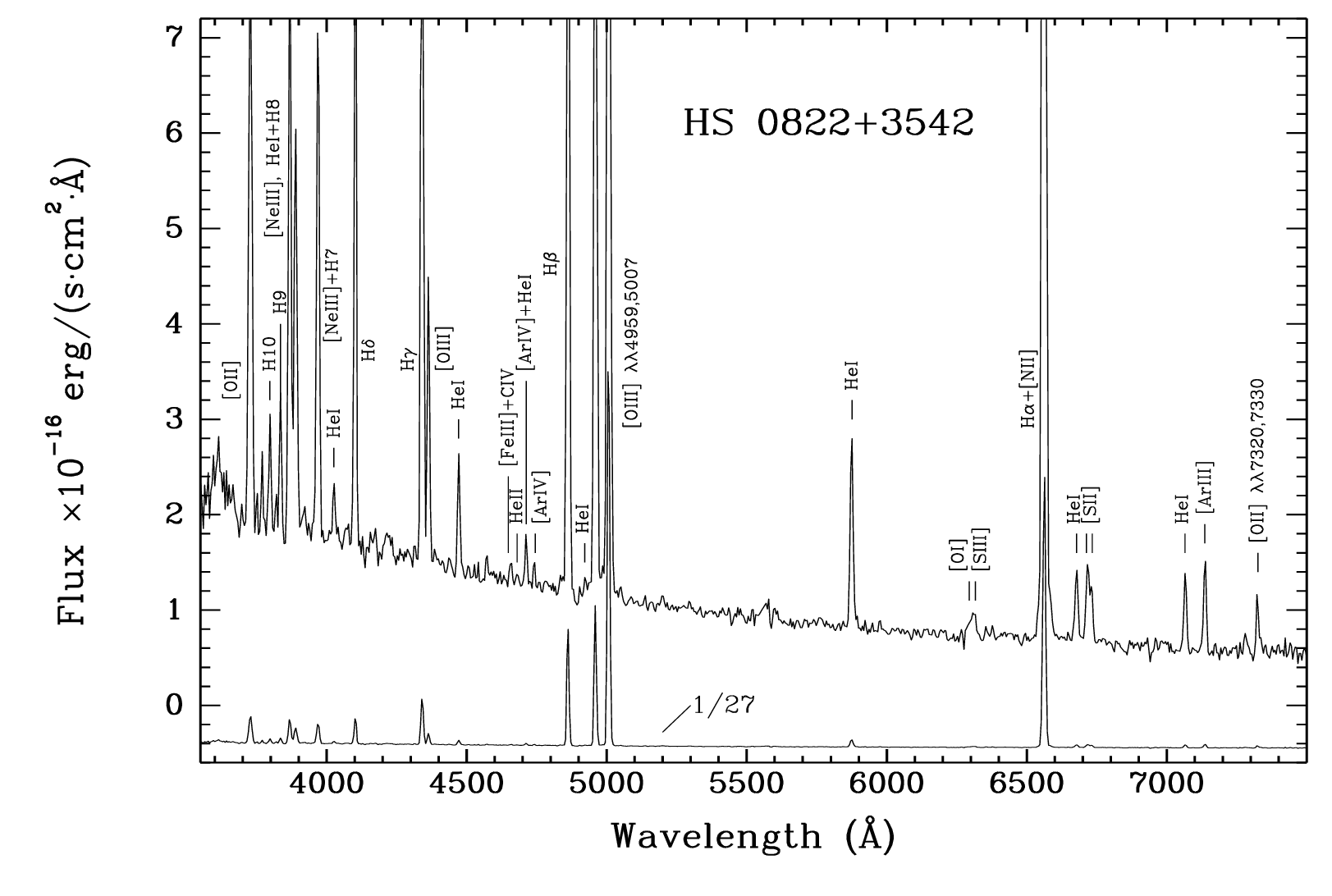}
      \caption{1D spectrum of HS~0822+3542 with a sampling of
	 4.6~\AA~pixel$^{-1}$.
	 The spectrum in the bottom is scaled by 1/27 to show the relative
         intensities of strong lines.
              }
         \label{fig:Spectrum_HS0822}
   \end{minipage}%
   \end{figure*}

The spectrum of the brightest knot of HS~0822+3542, extracted with an
aperture of 3\farcs6$\times$1\farcs2, is shown in Fig.
\ref{fig:Spectrum_HS0822}.
It is dominated by very strong emission lines. The relative intensities of all
emission lines, together with the equivalent width $EW$(H$\beta$), extinction
coefficient $C$(H$\beta$) and the $EW$ of Balmer absorption lines are given in
Table~\ref{t:Intens}. $C$(H$\beta$) was derived from the Balmer decrement
using the self-consistent method of Izotov et al.~(\cite{Izotov94}).
The derived value of $C$(H$\beta$)=0.0$\pm$0.09 is consistent, to within the
uncertainties, with a reddening of $E(B-V)$ = 0.047$\pm$0.007, expected from
foreground extinction in the Galaxy (Schlegel et al. ~\cite{Schlegel98}).

We analyzed chemical abundances and physical parameters with the method
described by Kniazev et al.~(\cite{Kniazev00}).
The measured electron temperatures and density, and derived chemical
abundances are presented in Table \ref{t:Chem}.
Our new data give a slightly
higher value of 12+$\log$(O/H) (7.44$\pm$0.06 versus 7.35$\pm$0.04 from the
NOT spectrum), but the difference is not significant. The main source of this
difference is the intensity of the [\ion{O}{iii}]~$\lambda$4363 line,
relative to H$\beta$. Its measured value is 0.104 on the 6\,m telescope
spectrum, in comparison to 0.123 on the NOT spectrum. Part of the derived
differences in O/H could be due to
the differences in observational conditions, resulting in the sampling of
slightly different regions. However, both results are in fact consistent to
within their uncertainties. The ratios of other heavy element abundances
(\ion{Ne}{}, \ion{S}{}, \ion{N}{}) to that of oxygen, derived from the NOT
and the 6\,m telescope data, are also similarly consistent. With the higher
resolution (1.2~\AA\ pixel$^{-1}$) spectrum near H$\alpha$ we improved the
precision
of the \ion{N}{} abundance, since the [\ion{N}{ii}]~$\lambda$6584 emission
line was detected with a higher S/N ratio than in the NOT spectrum.
The intensities of \ion{Ar}{} lines are measured for the first time in this
BCG,
and we present here its abundance. At the position of the [\ion{Fe}{iii}] line
$\lambda$4658~\AA\ we detected a signal at a level of 4$\sigma$.
However, if $\log$(Fe/O) is typical of other very metal-poor BCGs
($\sim -1.65$), the signal in this line should be only $\sim$1.3$\sigma$.
One possible explanation of the strength of this line is the contribution of
the WR spectral feature
\ion{C}{iv}~$\lambda$4658. In many BCGs with detected WR features the
intensity of the \ion{C}{iv}~$\lambda$4658 line is comparable to that of
\ion{He}{ii}~$\lambda$4686 (e.g., Guseva et al. \cite{Guseva00}). Thus, a
likely interpretation of the observed feature is that it is the sum of the
lines of [\ion{Fe}{iii}] and \ion{C}{iv}.
This suggests that higher S/N spectroscopy could
detect the other WR features in this young starburst region.
The \ion{He}{ii}~$\lambda$4686 line in our spectrum is broadened to
$\sim$15~\AA, but the S/N ratio is too low to accept this as direct
evidence of a WR population.

\begin{table}[tb]
\centering{
\caption{Line intensities of HS~0822+3542}
\label{t:Intens}
\begin{tabular}{lcc} \hline \hline
\rule{0pt}{10pt}
$\lambda_{0}$(\AA) Ion                  & $F$($\lambda$)/$F$(H$\beta$)&$I^a$($\lambda$)/$I$(H$\beta$) \\ \hline
3727\ [\ion{O}{ii}]\                      & 0.3116$\pm$0.0239 & 0.3061$\pm$0.0255 \\
3868\ [\ion{Ne}{iii}]\                    & 0.2381$\pm$0.0224 & 0.2339$\pm$0.0232 \\
3889\ \ion{He}{i}\ +\ H8\                 & 0.1599$\pm$0.0178 & 0.1846$\pm$0.0219 \\
3967\ [\ion{Ne}{iii}]\ +\ H7\             & 0.1966$\pm$0.0145 & 0.2217$\pm$0.0180 \\
4026\ \ion{He}{i}\                        & 0.0209$\pm$0.0036 & 0.0205$\pm$0.0036 \\
4101\ H$\delta$\                          & 0.2281$\pm$0.0170 & 0.2495$\pm$0.0199 \\ 
4340\ H$\gamma$\                          & 0.4413$\pm$0.0316 & 0.4567$\pm$0.0340 \\ 
4363\ [\ion{O}{iii}]\                     & 0.1044$\pm$0.0079 & 0.1026$\pm$0.0080 \\
4471\ \ion{He}{i}\                        & 0.0390$\pm$0.0036 & 0.0383$\pm$0.0036 \\
4658\ [\ion{Fe}{iii}] +\ \ion{C}{iv}\     & 0.0081$\pm$0.0021 & 0.0080$\pm$0.0021 \\
4686\ \ion{He}{ii}                        & 0.0053$\pm$0.0015 & 0.0052$\pm$0.0015 \\
4713\ [\ion{Ar}{iv]}\ +\ \ion{He}{i}\     & 0.0158$\pm$0.0031 & 0.0155$\pm$0.0031 \\
4740\ [\ion{Ar}{iv]}\                     & 0.0071$\pm$0.0021 & 0.0070$\pm$0.0021 \\
4861\ H$\beta$\                           & 1.0000$\pm$0.0720 & 1.0000$\pm$0.0734 \\ 
4959\ [\ion{O}{iii}]\                     & 1.2104$\pm$0.0877 & 1.1889$\pm$0.0877 \\
5007\ [\ion{O}{iii}]\                     & 3.6425$\pm$0.2637 & 3.5779$\pm$0.2639 \\
5876\ \ion{He}{i}\                        & 0.0852$\pm$0.0064 & 0.0837$\pm$0.0066 \\
6300\ [\ion{O}{i}]\                       & 0.0082$\pm$0.0029 & 0.0081$\pm$0.0029 \\
6312\ [\ion{S}{iii}]\                     & 0.0099$\pm$0.0028 & 0.0098$\pm$0.0028 \\
6563\ H$\alpha$\                          & 2.7842$\pm$0.1990 & 2.7510$\pm$0.2172 \\ 
6584\ [\ion{N}{ii}]\                      & 0.0105$\pm$0.0029 & 0.0103$\pm$0.0029 \\
6678\ \ion{He}{i}\                        & 0.0282$\pm$0.0038 & 0.0277$\pm$0.0039 \\
6717\ [\ion{S}{ii}]\                      & 0.0268$\pm$0.0037 & 0.0263$\pm$0.0038 \\
6731\ [\ion{S}{ii}]\                      & 0.0182$\pm$0.0034 & 0.0179$\pm$0.0034 \\
7065\ \ion{He}{i}\                        & 0.0168$\pm$0.0042 & 0.0165$\pm$0.0043 \\
7136\ [\ion{Ar}{iii}]\                    & 0.0244$\pm$0.0035 & 0.0239$\pm$0.0036 \\
  & & \\
$C$(H$\beta$)\ dex         & \MC {2}{c}{0.00$\pm$0.09} \\
$EW$(abs)\ \AA\            & \MC {2}{c}{4.90$\pm$0.67} \\
$EW$(H$\beta$)\ \AA\       & \MC {2}{c}{271$\pm$14}   \\
\hline \hline
\MC{3}{l}{$^a$ corrected for interstellar extinction and underlying}\\
\MC{3}{l}{~~~stellar absorption}\\
\end{tabular}
 }
\end{table}

\section{Discussion}
\label{Discussion}

\subsection{General parameters of the system}
\label{General}

In Table~\ref{t:Param} we present the main parameters relevant for further
discussion of the properties and status of these dwarf galaxies.
Their small mutual projected distance (3\farcm5, or $\sim$11 kpc) and
relative velocity ($\Delta V < 25$~\kms, Chengalur et al. \cite{GMRT03})
imply that both galaxies are physically associated.
Some of the parameters for HS~0822+3542 in Table~\ref{t:Param} have been
revised from those in Kniazev et al. (\cite{Kniazev00}). In particular,
the distance-dependent parameters have changed due to an improved distance
estimate. \ion{H}{i} related parameters have also changed due to
correction of the 21-cm line flux (GMRT, Chengalur et al. \cite{GMRT03}).
The integrated \ion{H}{i} flux presented by Kniazev et al.
(\cite{Kniazev00}), based on observations with the NRT, appeared to be off
by a factor of two due to the effects of confusion with the galaxy
SAO~0822+3545.
The adopted oxygen abundance is the weighted mean of current and previous
(Kniazev et al. \cite{Kniazev00}) values.

\begin{table}[tb]
\centering{
\caption{Abundances in HS~0822+3542}
\label{t:Chem}
\begin{tabular}{lc} \hline \hline
\rule{-3pt}{10pt}
$T_{\rm e}$(\ion{O}{iii})(K)\                & 18200$\pm$1000 ~~     \\
$T_{\rm e}$(\ion{O}{ii})(K)\                 & 15100$\pm$800 ~~     \\
$T_{\rm e}$(\ion{S}{iii})(K)\                & 16800$\pm$800 ~~     \\
$N_{\rm e}$(\ion{S}{ii})(cm$^{-3}$)\         & $\le$10 ~~         \\
& \\
O$^{+}$/H$^{+}$($\times$10$^5$)\     & 0.259$\pm$0.040~~     \\ 
O$^{++}$/H$^{+}$($\times$10$^5$)\    & 2.478$\pm$0.345~~     \\ 
O/H($\times$10$^5$)\                 & 2.737$\pm$0.348~~     \\ 
12+$\log$(O/H)\                         & ~7.44$\pm$0.06~~      \\
& \\ 
N$^{+}$/H$^{+}$($\times$10$^7$)\     & 0.753$\pm$0.178~~     \\ 
ICF(N)\                              & 10.566                 \\ 
$\log$(N/O)\                            & --1.54$\pm$0.12~~     \\
& \\ 
Ne$^{++}$/H$^{+}$($\times$10$^5$)\   & 0.338$\pm$0.053~~     \\ 
ICF(Ne)\                             & 1.105                 \\ 
$\log$(Ne/O)\                           & --0.87$\pm$0.09~~     \\
& \\ 
S$^{+}$/H$^{+}$($\times$10$^7$)\     & 0.430$\pm$0.059~~     \\ 
S$^{++}$/H$^{+}$($\times$10$^7$)\    & 3.569$\pm$1.091~~     \\ 
ICF(S)\                              & 2.542                 \\ 
$\log$(S/O)\                            & --1.43$\pm$0.13~~     \\
& \\ 
Ar$^{++}$/H$^{+}$($\times$10$^7$)\   & 0.713$\pm$0.114~~     \\ 
Ar$^{+++}$/H$^{+}$($\times$10$^7$)\  & 0.660$\pm$0.210~~     \\ 
ICF(Ar)\                             & 1.008                 \\ 
$\log$(Ar/O)\                           & --2.30$\pm$0.09~~     \\
\hline \hline
\end{tabular}
 }
\end{table}


\begin{table}
\caption{Main parameters of the studied galaxies}
\label{t:Param}
\begin{tabular}{lcc} \\ \hline \hline
Parameter                      & HS~0822+3542        & SAO~0822+3545    \\ \hline
R.A.(J2000.0)                  & ~~08 25 55.47       & ~~08 26 05.59     \\
DEC.(J2000.0)                  & $+$35 32 32.9       & $+$35 35 25.7     \\
A$_{\rm B}$ (from NED)      & 0.20                & 0.20              \\
B$_{\rm tot}$               & 17.92$\pm$0.05$^{(1)}$  & 17.56$\pm$0.03$^{(2)}$    \\
(B$-$V)$_{\rm tot}^0$       & 0.27$\pm$0.07$^{(1)}$   & 0.08$\pm$0.06$^{(2)}$ \\
(V$-$R)$_{\rm tot}^0$       & 0.14$\pm$0.09$^{(1)}$   & 0.14$\pm$0.06$^{(2)}$ \\
V$_{\rm hel}(\ion{H}{i})$ (\kms)   &  716$^{(8)}$          & 738$^{(8)}$  \\
Dist$^{(3)}$ (Mpc)               & 11.0                & 11.0            \\
M$_{\rm B}^0$ $^{(4)}$         &  --12.49            & --12.85          \\
Opt. size (\arcsec)$^{5}$      & 14.8$\times$7.4$^{(1)}$ & 28.2$\times$15.5$^{(2)}$  \\
Opt. size (kpc)                & 0.79$\times$0.39    & 1.50$\times$0.83$^{(2)}$  \\
12+$\log$(O/H)  \                 & 7.38$^{(1,2)}$          & ---               \\
\ion{H}{i} int.flux$^{(6)}$      & 0.34$^{(8)}$        & 0.98$^{(8)}$      \\
M(\ion{H}{i}) (10$^{7} M_{\odot}$)  & 0.97$^{(2)}$    & 2.8$^{(2)}$       \\
M(\ion{H}{i})/L$_{\rm B}$$^{(7)}$   & 0.63$^{(2)}$    & 1.31$^{(2)}$       \\   \hline\hline

\multicolumn{3}{l}{(1) -- from Kniazev et al. (\cite{Kniazev00}); (2) -- derived in this paper } \\
\multicolumn{3}{l}{(3) -- distance derived according to the kinematics of the}  \\
\multicolumn{3}{l}{~~~~~~~Local Volume (Karachentsev \& Makarov \cite{KM01})} \\
\multicolumn{3}{l}{(4) -- corrected for Galactic extinction A$_{\rm B}$} \\
\multicolumn{3}{l}{(5) -- $a \times b$ at $\mu_{\rm B}=$25\fm0~arcsec$^{-2}$} \\
\multicolumn{3}{l}{(6) -- in units of Jy$\cdot$\kms; (7) -- in units of ($M/L_\mathrm{B}$)$_{\odot}$}  \\
\multicolumn{3}{l}{(8) -- from Chengalur et al. (\cite{GMRT03})} \\
\end{tabular}
\end{table}

\subsection{On the evolutionary status of SAO~0822+3545}
\label{status}

\subsubsection{The very blue colours of SAO~0822+3545}

The integrated colours of SAO 0822+3545 are unusually blue. Only two out of
about 250 LSB/dIrr galaxies with known integrated colours $(B-V)$, $(V-R)$, or
$(B-R)$ (from papers by Ronnback \& Bergvall \cite{Ronnback94}, 
McGaugh \& Bothun \cite{McGaugh94}, 
de Blok et al. \cite{Blok95}, 
van Zee et al. \cite{Zee97}, 
O'Neil et al. \cite{O'Neil97}, 
van Zee et al. \cite{Zee01},  
and Burkholder et al. \cite{BIS01})
have such blue colours.
Only one of the 65 dIrr galaxies
studied by Makarova et al. (\cite{Makarova98}), 
Makarova \& Karachentsev (\cite{MK98}),  
and Makarova (\cite{Makarova99})  
appeared that blue; Makarova et al. (\cite{Makarova98}) noted that this
particular LSBG (UGCA~292) has
several blue stellar complexes, in which van Zee (\cite{Zee00}) detected
strong line emission, indicating young starbursts.

Thus, the unusual colours of SAO~0822+3545 could be due to its recent
enhanced SF. To estimate the age of its stellar population, we can compare
its $(B-V)^\mathrm{0}_\mathrm{tot}$ and $(V-R)^\mathrm{0}_\mathrm{tot}$
colours with model values. Unfortunately, due to the age-metallicity
degeneracy, similar colours can correspond to very different ages.
Therefore some \mbox{{\it a priori}} information on galaxy metallicity is
necessary
to disentangle the degeneracy. This can be obtained from the dIrr galaxy
metallicity--luminosity relation (Skillman et al. \cite{Skillman89}; Pilyugin
\cite{Pilyugin01}). For SAO~0822+3545 it predicts $Z\sim1/20~Z$\sunn.
In fact, for LSB galaxies this relation likely goes significantly below
(e.g., Kunth \& \"Ostlin \cite{Kunth2000}) that from Skillman et al.,
so 1/20~$Z$\sunn\ is probably the upper limit for the metallicity of
SAO~0822+3545.

\subsubsection{Comparison of observed and model parameters}
\label{text:tracks}

\begin{figure}
    \includegraphics[width=8.5cm]{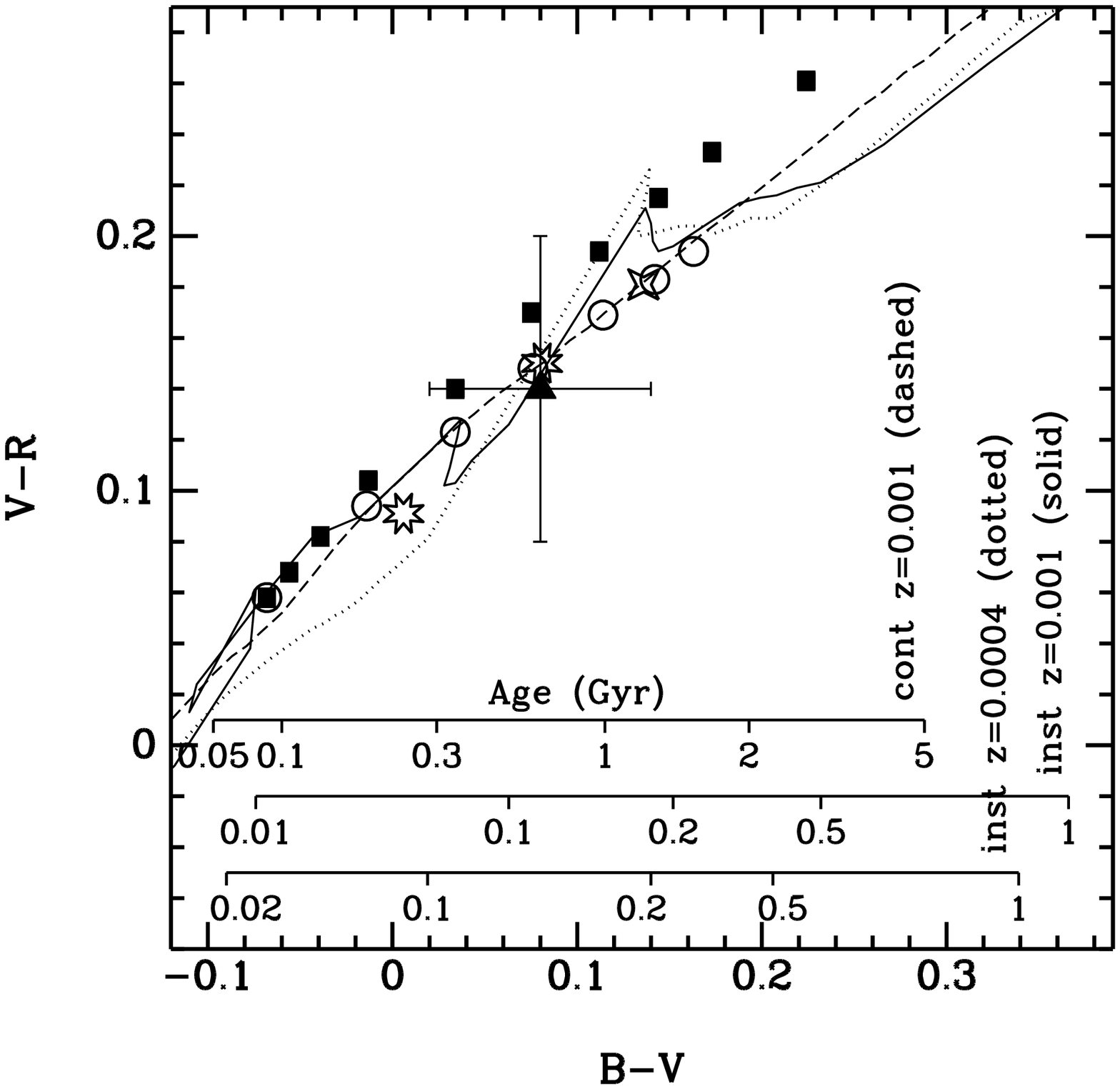}
    \caption{
      $BVR$ tracks based on PEGASE.2 evolution models from Fioc
      \& Rocca-Volmerange (\cite{FRV97}). Solid and dotted lines are for
      instantaneous starburst stellar populations with $Z$ of 1/20~$Z$\sunn\
      and 1/50~$Z$\sunn, respectively (Salpeter IMF, $M_{\rm up}$ and
      $M_{\rm low}$ are 120 and 0.1~$M$\sunn).
      The dashed line is for 1/20~$Z$\sunn\
      continuous SF at a constant rate. Time scales (in Gyr) for respective
      tracks are shown in the bottom. The filled triangle with error bars
      shows corrected for Galactic extinction of $E(B-V)=$~0.047 the total
      $BVR$ colours of SAO 0822+3545.
      Filled squares show the sequence of $BVR$ colors for composite stellar
      populations with a ``young'' (10 Myr) starburst and an admixture of a
      10~Gyr (``old'') coeval population. The mass ratio
      $\eta = M_{\rm old}/M_{\rm young}$ varies along the sequence
      as 0,2,5,10,20,30,40,50,60 and 80. Open circles show the colours of
      a composite population with the same 10~Myr ``young'' population mixed
      with a 250~Myr (``old'') coeval population. For this case the
      mass ratio $\eta$ sequence is shown for values 0,2,5,10,20,40 and 80.
      Two open stars show the $BVR$ colours of a stellar population
      resulting from continuous SF with a steep IMF ($\alpha$=3.35), lasting
      100 and 250 Myr, while the
       cross shows the colours for the case of continuous  SF, lasting
       1 Gyr, but for a Salpeter IMF with $M_{\rm up}$=30~$M$\sunn (see text).
      }
    \label{tracks}
   \end{figure}

To get some insight on the evolution status of SAO 0822+3545,
we compared its observed colours and $EW$(H$\alpha$)
with model predictions. For colours we used results derived
from PEGASE.2 models (Fioc \& Rocca-Volmerange \cite{FRV97, Pegase2}).
We
also used these models to estimate the mass of the stellar population.
We calculated spectral energy distributions (SEDs) for instantaneous SF
bursts with $Z$=1/20 and 1/50~$Z$\sunn, as well as the time
behavior of $B$, $V$, $R$ luminosities and the respective colour tracks.
In Fig.~\ref{tracks}, we show
$BVR$ colour tracks for instantaneous SF bursts with $Z$=1/20 and
1/50~$Z$\sunn, using solid and dotted lines,  respectively,
assuming a Salpeter IMF with
$M_{\rm low}=0.1~M$\sunn, $M_{\rm up}=120~M$\sunn.
The track for continuous SF with constant SFR ($Z=1/20~Z$\sunn)
and the same IMF is shown by dashed line. SAO~0822+3545 extinction
corrected  colours
(with their $\pm1\sigma$ uncertainties) are also plotted in the figure
as the filled triangle.

The $BVR$ colours of SAO~0822+3545
($(B-V)_0=$0.08$\pm$0.06, $(V-R)_0=$0.14$\pm$0.06) fall very close
to the model tracks. For
instantaneous SF with $Z=1/20~Z$\sunn\ they correspond to an age of
$\sim$110($\pm^{50}_{80}$) Myr. For continuous SF with a constant rate the
respective age is $\sim$0.6($\pm^{0.7}_{0.3}$) Gyr.
Thus, the integrated colours of this galaxy allow various
interpretations of its evolutionary status.

To further constrain the range of the models, consistent with the very
blue $BVR$ colours of SAO 0822+3545 we use two additional parameters:
the observed EW(H$\alpha$) and the dynamic mass of this galaxy.
To model the observed $EW$(H$\alpha$) we did not use PEGASE.2, since
a coherent computation of stellar and nebular emission would require coupling
with the CLOUDY code (Ferland \cite{Ferland96}; Moy et al. \cite{Moy01}).
Instead,
we compared values of $EW$(H$\alpha$) with the predictions of
Starburst99 (SB99, Leitherer et al. \cite{Starburst99}).
In Table \ref{Tab_models} we summarize all the models we have examined
in order to match both the observed $BVR$ and EW(H$\alpha$). Each considered
model also predicts a total stellar mass, which should be compatible with
$M_{\rm dyn}$. Since LSBGs are known as gas-rich galaxies, it is
worthwhile to compare this stellar mass with the total gas mass $M_{\rm gas}$
(see the estimates of $M_{\rm dyn}$ and $M_{\rm gas}$ in Section
\ref{gasmass}).

We will only briefly discuss the models from Table \ref{Tab_models} which
fail to match the three observational parameters, and then turn to a couple
of more realistic models.
First, simple scenarios with constant SFR and ages in the range of
$0.4-1.1$ Gyr, for the standard Salpeter IMF (Models 4 and 5, as the
representatives of the whole range of ages), should be excluded, since
the predicted $EW$(H$\alpha$) is much higher then the observed value
of 15 to 20 \AA.
Similarly, Model 3, with the same IMF, an instantaneous SF episode, and
an age of $\sim$100 Myr, is excluded due to the very small expected
$EW$(H$\alpha$).

Models 6 and 7 deal with IMFs biased to the low-mass end.
For a continuous SF law, lasting $\sim$250 and $\sim$100 Myr, respectively
(shown in Figure \ref{tracks} by empty stars), they have $BVR$ colours
consistent either with those observed, or with colours corrected for a
possible extinction $E(B-V)$ $\sim$0.10.
EW(H$\alpha$), according to SB99, appears to be consistent with the
observed values.
However, the mass of formed stars $M_{\rm star}$, as estimated with PEGASE.2,
is several times larger than the LSBG dynamic mass (see section
\ref{gasmass}).
Therefore both Model 6 and 7 should be rejected.

More realistic models include composite stellar populations created
by instantaneous SF episodes with the standard Salpeter IMF.
Indeed, since a single-component Model 3 fails to reproduce the observed
EW(H$\alpha$), we would need a younger Single Stellar (coeval) Population
(SSP), to give enough ionizing photons. But this would appear too blue,
and in order to reproduce the integrated $BVR$ colours of SAO 0822+3545,
one would need to prepare a mixture with many older redder stars.
For illustration, we show in Fig.~\ref{tracks} the colours of a series of
composite models with filled squares.
They consist  of various mixtures of two coeval populations (SSP) with ages of
$t_\mathrm{1}=10$~Myr and $t_\mathrm{2}=10$~Gyr.
Their mass ratio $\eta$ varies as explained in the figure caption.
Model 1 in Table \ref{Tab_models} with $\eta$=30 best matches both $BVR$
colours and EW(H$\alpha$), and its $M_{\rm star}$ is compatible with
$M_{\rm dyn}$ and $M_{\rm gas}$. The third SSP component (in parentheses)
with an age of $\sim$100 Myr and a mass comparable to that of the 10-Myr old
component could also be present, however it is not necessary for a good fit
to the observed data.

Another series of composite models is shown in Fig.~\ref{tracks} by open
circles. These models consist of two coeval populations with ages of
$t_\mathrm{1}=10$~Myr and $t_\mathrm{2}=250$~Myr. The model with $\eta$=10,
which best matches the observed $BVR$ and EW(H$\alpha$), is Model 2 in
Table \ref{Tab_models}. Again, $M_{\rm star}$ for this model is consistent
with $M_{\rm dyn}$ and $M_{\rm gas}$.

Such types of composite models with two coeval populations, one of which
has an age $t_\mathrm{1}=10$~Myr, can be constructed for any
$t_\mathrm{2}$ in the range between $\sim$250~Myr and $\sim$10 Gyr, and
$\eta$ between $\sim$10 and $\sim$30.
All of them provide good matches to the observed colours and EW(H$\alpha$),
and have stellar masses compatible with the dynamic and gas mass of
the LSBD.

Finally, we discuss one more model, with an IMF deficit in massive
stars relative to the `standard' IMF used in the discussion above.
This IMF has $\alpha$=2.35, and $M_{\rm up}$ and $M_{\rm low}$ of 30 and 0.1
$M$\sunn, respectively.
According to SB99 (Fig.84), a model with this IMF and continuous SF
(Model 8) will reach the observed value of EW(H$\alpha$)=20~\AA\ in
$\sim$1~Gyr.
The PEGASE.2 colours of such a model are $(B-V)$=0.14 and $(V-R)$=0.20 (open
cross in Fig.~\ref{tracks}). They are compatible with the observed colours
to within their uncertainties. The resulting $M_{\rm star}$ is also
compatible with the LSBG $M_{\rm dyn}$ and $M_{\rm gas}$.

Summarizing this point, we conclude that for a Salpeter IMF
with $M_{\rm up}$=120 $M$\sunn\
only a model with a composite population can explain both the photometry and
the $EW$(H$\alpha$) data.
It has to include a ``young'' coeval population with an
age of $\sim$10 Myr and an ``old'' coeval component. If the age of the
``old'' population is 10~Gyr, it should be $\sim$30 times more massive than
the ``young'' component; an ``old'' component with  $T \sim$250 Myr
should be 10 times more massive than the ``young'' one.
The real situation is likely somewhere in between these two extreme
cases.
In any case, for an ``old'' component with an age between 0.25 and
10 Gyr, its mass should be between 10 and 30 times the masses of the
``young'' population.
An ``intermediate'' age ($\sim$100~Myr) coeval population, having
model $BVR$ colours close to those observed for SAO 0822+3545, could also
contribute appreciable mass.

Some uncertainties of these estimates are related to the respective
uncertainties of the LSBD metallicity and/or uncorrected internal
extinction.
The variation of the former parameter by a factor of two will slightly change
the relative mass of "young" and "old" stellar components.
The internal extinction is usually small in LSB galaxies.
For example, the galaxy corrected $BVR$ colours of the LSBD would still lie
close to the model evolution tracks even with a correction for an internal
extinction of $E(B-V)$=0.10.
However, these colours would correspond to
a significantly smaller age, $\sim$30 Myr. The self-consistent
interpretation of the galaxy colours and observed $EW$(H$\alpha$) still
requires a composite model, similar to those discussed above. The main
effect of such reddening (if  present)  would be the significant
reduction (by a factor of 4--5) of the mass ratio, $\eta$.

For the IMF with a deficit of massive stars, both EW(H$\alpha$) and
$BVR$ colours can be explained by a single component with a constant SFR and
an age of $\sim$1 Gyr. The formed stellar mass would be consistent with
$M_{\rm dyn}$. Note, however, that such a model would be inconsistent with
the recent encounter of the LSBD and BCG, and its resulting disturbance
(section \ref{interaction}).
In addition, internal extinction on the level of
$E(B-V)$=0.10 would cause problems for such a model, since this case
would require ages of only 100--250 Myr, an EW(H$\alpha$) that is too large
($\gtrsim$50~\AA), and $M_{\rm star} \sim M_{\rm dyn}$.

The available data on the integrated $BVR$-colours and
$EW$(H$\alpha$) in SAO 0822+3545 imply that the significant or main stellar
mass fraction is related to young populations with ages from 10 Myr to
1 Gyr. In case of a `standard' Salpeter IMF with $M_{\rm up}$=120 $M$\sunn,
this LSBG likely experienced at least two localized SF episodes.
Continuous SF, lasting $\sim$1 Gyr, can match all
available data with a single population, if the IMF is significantly biased
to the low-mass end. To distinguish which of the two SF scenarios
is more probable, we would need to measure the integrated $BVR$ colours of
SAO 0822+3545 with better precision ($\sigma \lesssim$0\fm02), and determine
the amount of internal extinction.


\begin{table*}
\begin{center}
\caption{Parameters of various evolution models vs those observed in SAO 0822+3545}
\label{Tab_models}
\begin{tabular}{lllcccc} \\ \hline \hline
\MC{1}{c}{Model IMF}     &
\MC{1}{c}{Model Nu.}     &
\MC{1}{c}{ SF type}       &
\MC{1}{c}{ Mass }   &
\MC{1}{c}{ $(B-V)^{0}$ } &
\MC{1}{c}{ $(V-R)^{0}$ } &
\MC{1}{c}{ EW(H$\alpha$) } \\

\MC{1}{c}{ }       &
\MC{1}{c}{ }       &
\MC{1}{c}{\& Age }       &
\MC{1}{c}{(in $M$\sunn) }    &
\MC{1}{c}{  } &
\MC{1}{c}{  } &
\MC{1}{c}{(in \AA) }   \\

\MC{1}{c}{ (1) } &
\MC{1}{c}{ (2) } &
\MC{1}{c}{ (3) } &
\MC{1}{c}{ (4) } &
\MC{1}{c}{ (5) } &
\MC{1}{c}{ (6) } &
\MC{1}{c}{ (7) } \\
\hline
\\[-0.3cm]
$\alpha$=2.35    & 1&SSP$_{1}$ $t_{1}$=10~Myr    & 4.4 10$^5$  &  0.07    & 0.17  &  20   \\
0.1--120~$M$\sunn&  &SSP$_{2}$ $t_{2}$=10~Gyr    & 1.3 10$^7$  &          &       &       \\
		 &  &(SSP$_{3}$ $t_{3}$=100~Myr) &(4.4 10$^5$) &          &       &       \\
		 &  \MC{6}{l}{------------------------------------------------------------------------------------------------------ } \\

		 & 2&SSP$_{1}$ $t_{1}$=10~Myr    & 2.4 10$^5$  &  0.07    & 0.15  &  20   \\
		 &  &SSP$_{2}$ $t_{2}$=250~Myr   & 2.4 10$^6$  &          &       &       \\
		 &  \MC{6}{l}{------------------------------------------------------------------------------------------------------ } \\
		 & 3&SSP$_{1}$  $t_{1}$=100~Myr  & 2.5 10$^6$  &  0.06    & 0.13  &$<$0.1 \\
		 &  \MC{6}{l}{------------------------------------------------------------------------------------------------------ } \\
		 & 4&Con$_{1}$ $t_{1}$=0.4~Gyr   & 1.2 10$^8$  &  0.04    & 0.12  & 100   \\
		 & \MC{6}{l}{------------------------------------------------------------------------------------------------------- } \\
		 & 5&Con$_{2}$ $t_{2}$=1.1~Gyr   & 7.8 10$^7$  &  0.13    & 0.20  & 150   \\
\hline
\\[-0.3cm]
$\alpha$=3.35    & 6&Con$_{1}$ $t_{1}$=250~Myr   & 2.0 10$^9$  &  0.08    & 0.15  &  20   \\
0.1--120~$M$\sunn& \MC{6}{l}{------------------------------------------------------------------------------------------------------- } \\
		 & 7&Con$_{2}$ $t_{2}$=100~Myr   & 4.3 10$^9$  &  0.01    & 0.09  &  20$^a$   \\

\hline
\\[-0.3cm]
$\alpha$=2.35    & 8&Con$_{1}$ $t_{1}$=1~Gyr     & 8.7 10$^7$  &  0.14    & 0.18  &  20   \\
0.1--30~$M$\sunn &  &                            &             &          &       &       \\
\hline \hline \\[-0.2cm]
Observed in      &  & ~~~~~~~~~~~~~~$M_{\rm gas}$  = & 3.7 10$^7$  &  0.08    & 0.14    &  20   \\
SAO 0822+3545    &  & ~~~~~~~~~~~~~~$M_{\rm dyn}$  = & 4--7 10$^8$  &$\pm$0.06 &$\pm$0.06&       \\
\hline \hline \\[-0.2cm]
\MC{7}{l}{$^a$ $BVR$ colours correspond to those observed, after correction for possible internal}\\
\MC{7}{l}{~~~~~~extinction of $E(B-V)$=0.1.}  \\
\end{tabular}
\end{center}
\end{table*}

\subsubsection{The LSBD gas mass fraction}
\label{gasmass}

The ratio of gas and stellar masses is one of the important parameters
characterizing the evolutionary status of the galaxy.
The gas-mass fraction  (relative to the full visible mass) of several XMD BCGs
reaches values of $0.94-0.98$. This is a good indication of their
possible youth. In addition, Schombert et al. (\cite{SME01}), in a
recent
study of a large sample of LSBGs, discovered several galaxies with
gas-mass fractions of 0.8--0.85. Perhaps these are the galaxies with the
lowest known SFRs and/or relatively young objects.
To estimate the masses of various populations in SAO 0822+3542 we used
its $B$-band luminosity. We also assumed $Z=1/20~Z$\sunn\  and
considered the PEGASE.2 models, presented in Table \ref{Tab_models}.

The full gas mass of the LSBD is derived directly from \ion{H}{i} data in
Table \ref{t:Param}, accounting for a helium mass-fraction of 0.24:
$M_{\rm gas}\sim$3.7$\times$10$^7~M$\sunn. Then, for `Model 1' (10 Gyr
``old'' stellar population), the gas mass fraction is
$f_{\rm g} \equiv M_{\rm gas}$/($M_{\rm gas} + M_{\rm star}$) = 0.73.
For `Model 2' (the ``old'' stellar population 0.25-Gyr old), the gas mass
fraction is $f_{\rm g}$=0.93.
For the LSBD ``old'' coeval population with an age between 0.25 and 10 Gyr,
the gas mass fraction will thus fall in the range of 0.73 to 0.93.
For `Model 8' with continuous SF, the gas mass fraction would be only
$\sim$0.3.
While the most probable SF scenario for this LSBD remains somewhat uncertain,
it could be  one of the most gas-rich galaxies of its type.

One more parameter, important for comparison with models, is the total dynamic
mass of galaxy. From the NRT \ion{H}{i} profile of Kniazev et al.
(\cite{Kniazev00}), in which the LSBD contributes more than half of the
flux, we get an upper limit of $V_{\rm rot}$ of 20~\kms. Corrected for an
inclination angle of 58.5\degr\ (subsection \ref{Morphology}) this yields
$V_{\rm rot}$ of 23.5~\kms.
For the commonly assumed gas disk radius of 3--4 times $R_{\rm opt}$, this
results in $M_{\rm dyn}$=(4--7)$\times$10$^{8}~M$\sunn.

\subsection{Global environment: The Lynx-Cancer Void}
\label{minivoid}

\begin{figure*}[tbh]

   \begin{minipage}[t]{\linewidth}
   \centering
   \includegraphics[angle=-90,width=16cm]{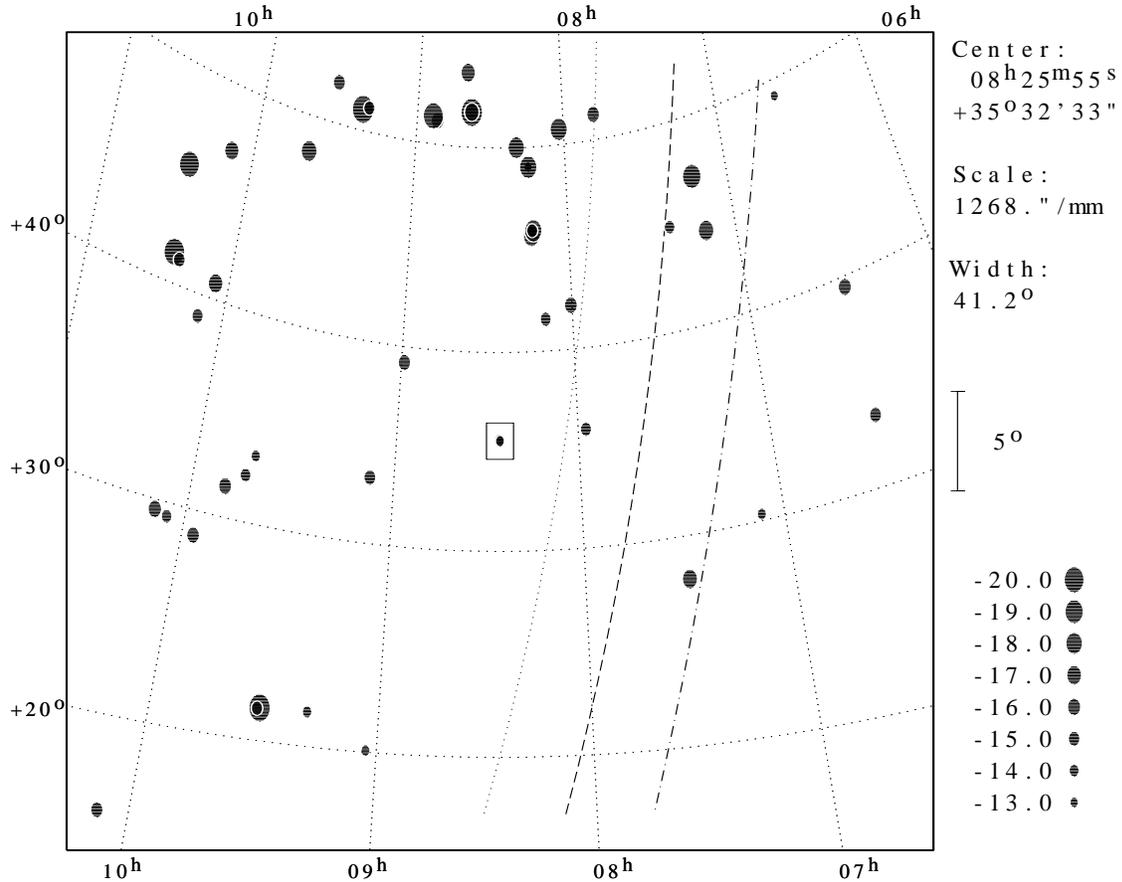}
    \caption{The position of HS 0822+3542 and SAO 0822+3545 (in the box) is
    shown
    relative to the other galaxies from LEDA in the redshift range between
    $V_{\rm hel} =$ 427 and 1027~\kms. The blue luminosities of galaxies are
    indicated by the size of the filled circle. The nearest neighbour to the
    pair,
    at the position R.A.$\sim8^h$, DEC.$\sim$36\degr\ ($D=$1.15 Mpc) has
    $M_{\rm B}= -$14.7. Dotted, dashed and dot-dashed lines show
    the projection of galactic latitudes $b^{II} =$30\degr, 25\degr\ and
    20\degr, respectively. Galactic extinction should more strongly affect
    the detection of faint galaxies. The absence of bright galaxies is not a
    selection effect, since some of them are found in the zone of
    $b^{II} < 25$\degr,
    }
         \label{void}
   \end{minipage}%
   \end{figure*}

HS~0822+3542 is well-isolated with regard to known, sufficiently bright
galaxies.
There are no cataloged galaxies in the NED and LEDA databases with
$V_{\rm hel} <$ 1200~\kms\ and angular distances less than 5.8\degr.
All the nearest cataloged galaxies (at projected distances of 1.15--1.45
Mpc) are dwarfs with $M_{\rm B}$ in the range --13.5 to --16.6. The
position of HS~0822+3542 and SAO~0822+3545 is shown (in the box) in Fig.
\ref{void}. The positions of other cataloged galaxies with radial velocities
of $V_{\rm hel}=V_{\rm pair}\pm300$~\kms\ ($\pm$4~Mpc) are shown by
filled circles, where $V_{\rm pair}=727$~\kms\ is the mean velocity of the
pair. Absolute magnitudes of all galaxies in the plot (corrected for
Galactic extinction according to Schlegel et al. \cite{Schlegel98}) are derived
from their cataloged magnitudes. The corrections for Local Group motion,
and the value of the local Hubble constant (67.5~\kms~Mpc$^{-1}$) in this
direction were made according to Karachentsev \& Makarov (\cite{KM01}).

There are six bright galaxies ($M < M_{\rm B}^{*}=-$19.73) in the plot
area. The latter corresponds to a volume element with sides (at $D=$~11
Mpc) of 9.5 Mpc $\times$7.8 Mpc (R.A.$\times$Dec.), and $\sim$9~Mpc in depth.
All of them
(shown by the largest filled circles) are situated at the
three-dimensional  distances
from the dwarf pair greater than 3.0~Mpc. Seven intermediate luminosity
galaxies with
$-19.73 \le M_{\rm B} < -$17.73 (shown by slightly smaller
circles), are also situated along the periphery of this volume. Only a few
real dwarfs are found inside this very large region.
With regard to bright galaxies, the pair of dwarfs in question is situated in
a void, not far from the void center. The void depth in radial velocity
through its center is of order
$\sim$800~\kms\ (from 400 to 1200 \kms, or $\sim$12 Mpc).
This is comparable to its
projected extent on the sky. This void is similar to the other nearby voids
discussed by Fairall (\cite{Fairall98}). We have named it the Lynx-Cancer
void, for the constellations on which this void is projected.
The parameters of this and the other similar voids are presented in
Table \ref{tab:voids}.
The latter are taken from Fairall's (\cite{Fairall98}) Table 4.1.
The first column gives the void's designation. In columns 2, 3 and 4 the
radial velocity extent, and the approximate equatorial coordinates of the
void's center are given.
In column 5 we present the void's central heliocentric velocity. Columns 6 and 7
give the galactic longitude and latitude of the void's center. Finally,
columns  8, 9 and 10 present the supergalactic coordinates.


\begin{table*}
\caption{Nearby voids parameters}
\label{tab:voids}
\begin{tabular}{lrrrrrrrrr} \\ \hline \hline
Designation& Size & RA    & Dec  & $cz$ &l$^{II}$& b$^{II}$&SGX&SGY &SGZ  \\
	   & \kms & hour  & \degr& \kms & \degr  & \degr  &\kms&\kms&\kms \\ \hline
Cetus      & 500  &  2.0  & -20  & 700  & 192    &$-72$   &100 &-600&-200 \\
Cepheus    & 500  & 23.5  & +65  & 800  & 112    &   5    &700 &   0&300  \\
Crater     & 500  & 11.5  & -15  & 1500 & 126    &$-28$   &1300&-700&200  \\
Volans     & 700  &  7.0  & -70  & 800  & 281    &$-25$   &-600&-300&-500 \\
Lynx-Cancer& 800  &  8.0  & +30  & 800  & 191    &  27    & 400& 400&-500 \\
Monoceros  &1000  &  4.0  & +05  & 800  & 185    &$-34$   & 400&-400&-500 \\ \hline
\end{tabular}
\end{table*}

Understanding the properties of galaxy populations in voids is
important for both galaxy and structure formation models.
It is natural to expect that candidate truly young galaxies would be very
isolated
(e.g., Peebles \cite{Peebles01}). Observationally, however, the situation
is unclear due to very poor statistics. For example, SBS 0335--052, one of
the best
candidate young galaxies, is situated at the outskirts of a loose
galaxy group.
It is interesting to note, however, that the recent data indicate that
at least several new extremely
metal-deficient (XMD) BCGs ($Z \lesssim 1/20~Z$\sunn), including HS~0822+3542,
and  HS~2236+1344 (Ugryumov et al. \cite{HSS-LM}), indeed lie in
regions of very low (massive) galaxy density. A larger sample of XMD BCGs
would be helpful for determining whether they really favour a void-type
environment.

\subsection{Supershell and young clusters in HS~0822+3542}

From the parameters of the supershell, we can estimate its characteristic age,
and thus, the time elapsed since the SF episode which produced this
supershell.
The relation between the shell age, its velocity and radius, for the case in
which the radiative cooling of hot gas inside the shell can be neglected
(readily
applicable for multiple SN-blown bubbles), is given as
$t_{\rm shell} = 0.6\times R/V_{\rm shell}$ (e.g., Weaver et al.
\cite{Weaver77}; Mac Low \& McCray \cite{MM88}).
This relation also holds approximately for the thin
shell phase, in which a bubble is pumped by a stationary stellar wind.
The observed values are $V_{\rm obs}=$30~\kms\ and $R=$0.24 kpc.
Both the maximal velocity and the apparent size of the shell in a simple
case depend similarly on the inclination angle. Thus, no inclination
correction is necessary. The shell age is then $t_{\rm shell}\sim$5 Myr.

The age of the current SF burst can be estimated from the observed
$EW$(H$\beta$)=271~\AA\ (see Table \ref{t:Intens}) using the results of
instantaneous SF burst models (Leitherer et al. \cite{Starburst99}). For a
Salpeter IMF with $M_{\rm low}$ and $M_{\rm up}$ of 1 and 100 $M$\sunn,
respectively, this value of $EW$(H$\beta$) corresponds to a starburst age
of $\sim$3~Myr. It is important to note that both the main part of the
starburst
continuum and the highly concentrated H$\beta$ emission of the corresponding
\ion{H}{ii} region were extracted for the analyzed spectrum.

Hence, the expected corrections for
$EW$(H$\beta$) are insignificant, and the estimate of starburst age is
reliable. It is also worth noting that, for the Salpeter IMF models with
$M_{\rm low}$=0.1~$M$\sunn, the maximum value of $EW$(H$\beta$) is
$\sim$230~\AA, and, hence, these models
cannot reproduce such a high $EW$(H$\beta$);
this could be due to a ``top-heavy'' IMF in this object.

For further analysis, we assume, that SN explosions turn on in
$\sim$3.5~Myr after the starburst beginning (e.g., Leitherer
\cite{Leitherer94}; Leitherer \& Heckman \cite{LH95}, Fioc \&
Rocca-Volmerange \cite{FRV97}).
The energy injected by stellar winds should also be considered when
calculating shell parameters.
For example, for $Z=Z$\sunn\ the kinetic energies injected by winds and
by SNe during the first 5 Myr are comparable. However, the former
scales as $Z^{+1}$, and for  HS~0822+3542 ($Z \sim$1/30~$Z$\sunn)
the energy injected by winds will be much lower than that of SNe.
Accounting for SNe delay and the shell age, we conclude
that the active SF in this region of HS~0822+3542 has lasted at least
8--9~Myr. Hence, the current SF episode is extended, probably with
propagating SF.

The shell kinetic energy, $E_{\rm kin}$, can be estimated simply from its
radius and velocity, suggesting some reasonable value of swept gas density.
For $V_{\rm shell}$=30~\kms, $R$=0.24 kpc and an average density of the
ambient gas of 0.1 cm$^{-3}$,  $E_{\rm kin}$ will be as following:
$E_{\rm kin}=0.5~M\cdot V^{2}=1.3\cdot10^{51}$ erg.
Accounting for the small efficiency of SN energy transfer to the
kinetic energy of the shell,  $\sim$0.1 (e.g., Thornton et al.
\cite{Thornton98}), the total energy of multiple SNe pumping the
observed BCG supershell is $\sim$1.3$\cdot10^{52}$ erg. This corresponds to
the energy of 13 typical SNe.
Taking the dynamical age of the supershell to be 5 Myr, we get a
SNe rate of 2.6 SNe~Myr$^{-1}$. For a Salpeter IMF with
$M_{\rm low}$ and $M_{\rm up}$, respectively, of 0.1 and 120 $M$\sunn,
from comparison with PEGASE.2 models, translates this to a total
coeval star cluster mass of $\sim$1.0$\cdot$10$^{4}M$\sunn.
The $M_{\rm B}$ of such a cluster with an age of 8.5
Myr would be --8.8. At the distance of
HS 0822+3542, the apparent $B$ magnitude of this cluster should be $B$=21.6.
The latter is $\sim$2\fm2 (or a factor of 7)
fainter than the estimate of $B$=19.46 found for
the current starburst component of HS~0822+3542 by Kniazev et al.
(\cite{Kniazev00}, Table 4, column 8). However, this value is quite consistent
with the intensity of knot ``b'', seen in the deconvolved image of
HS~0822+3542 just $\sim$1.3\arcsec\ to the
north of the main starburst region ``a''. Recall that its intensity is
estimated as $\sim$1/8 of that for knot ``a'' in subsection~\ref{curve}.
Thus, knot ``b'' could be the star cluster responsible for the
observed ionized gas supershell. The younger star cluster (knot ``a'')
is potentially the source of gas ionization in the shell. Its mass is
estimated
from $B_{\rm burst}=$19.46, corresponding to $M_{\rm B}^0=-$10.95.
From the PEGASE.2 model with the same IMF as above, a 3-Myr old starburst
with $M_{\rm B}^0=-$10.95 would have the mass of 3.9$\cdot$10$^4~M$\sunn.

It is possible that the shell age is slightly overestimated, while the
age of the young starburst (3 Myr) could be underestimated, so that they
could be more or less consistent with each other. In this case it is worth
considering the alternative option: that the
shell is pumped by this starburst. Then only the energy of stellar winds is
available.
According to Weaver et al. (\cite{Weaver77}), the kinetic energy
of the shell comprises only about 20\% of the total wind energy released
during its life-time. This implies a required wind energy of
6.5$\times$10$^{51}$~erg.
The temporal behavior of stellar wind kinetic luminosity is given in SB99.
As an example, this behavior is illustrated in Fig.13
of Efremov et al. (\cite{EPK02}) for the case of a 10$^6$~$M$\sunn\ young star
cluster with $M_{\rm low}=1~M$\sunn\ and $Z$=$Z$\sunn.
The estimate of the full wind kinetic energy released by such cluster
during the first 3 Myr is 9.9$\times$10$^{53}$ erg.
For $M_{\rm low}=0.1~M$\sunn, this energy is scaled down by a factor of 2.5.
The wind power scales with metallicity as $Z^{+1}$. Thus, for the case of
HS 0822+3542, that wind energy, scaled down by a factor of 30, yields a
value of 3.3$\times$10$^{52}$ erg (or 1.3$\times$10$^{52}$ erg for
$M_{\rm low}=0.1~M$\sunn).
That is, to provide the above required wind kinetic energy one would need
a 3-Myr old star cluster with a full mass (depending on $M_{\rm low}$)
of (2--5)$\times$10$^{5}$~$M$\sunn\  and $Z$=1/30~$Z$\sunn.
Then, with the PEGASE.2 model for $Z$=1/20~$Z$\sunn\ (the nearest
to that of HS 0822+3542) we find that such a star cluster would have
the blue luminosity, corresponding to $M_{\rm B}$=0.52 per 1 $M$\sunn.
Respectively, for masses of (2--5)$\times$10$^{5}$~$M$\sunn\  the cluster
expected absolute magnitudes are in the range of
$M_{\rm B}$ = --12.7 to --13.7. These are significantly brighter than the
observed $M_{\rm B}$=--10.95 of the current young starburst. Therefore,
this option does not look plausible.

We also revise the apparent stellar mass of the underlying ``disk''.
Based on the published value $B_{\rm disk}=$18.22 (Kniazev et al.
\cite{Kniazev00}), its
$M_{\rm B}^0$ is --12.19. According to the same PEGASE.2
model, for a starburst with an age of 100 Myr we have $M_{\rm B}$=
3.13 per 1~$M$\sunn. The ``disk'' stellar mass is then
$\sim$1.3$\cdot$10$^6~M$\sunn. Finally, we derive the BCG gas mass-fraction,
$f_{\rm g}$=0.90, using for $M$(HI) from Table \ref{t:Param} and a helium
mass fraction of 0.24.  The significant difference from the value in
Kniazev et al. (\cite{Kniazev00}) is a result of the reduced $M$(HI), as
explained in Section~\ref{General}.

\subsection{Triggering  mechanisms of SF activity of the pair}
\label{interaction}

The SF history of LSB galaxies is a complicated issue. The suggested options
range from SF episodes induced by distant encounters (O'Neil et
al. \cite{O'Neil98}) to low-level sporadic SF,
migrating across the galaxy body. The latter case can be considered as
a continuous process with the time-scale on the order of a galaxy's life-time
(Gerritsen \& de Blok \cite{GdB99}).
Since the pair of dwarf galaxies discussed here is situated deeply in
a void,
any interactions with massive galaxies on a cosmological timescale
would be very unlikely.

On the other hand,
the importance of interactions with low-mass neighbours to trigger
starbursts in BCG/\ion{H}{ii}-galaxies has been noted,
e.g., by Taylor et al. (\cite{Taylor95}), Pustilnik et al. (\cite{PKLU}),
and Noeske et al. (\cite{Noeske01}). The example of the BCG HS~0822+3542
indicates that, even in voids, SF bursts can be
triggered by the same mechanism (see also Pustilnik et al. \cite{HI_void}).

The projected distance between the dwarfs, 11.4 kpc, and their relative
radial velocity,
$\Delta V \le$ 25~\kms\ (Chengalur et al. \cite{GMRT03}), are both very small.
This implies rather slow collision, which is very efficient in exerting
mutual tidal forces.
The N-body simulation of a non-merging encounter between LSB and HSB (high
surface brightness) disk galaxies presented by Mihos et al. (\cite{Mihos97})
shows qualitatively different effects in each of the two galaxy types.
Due to the lower surface density of baryonic matter and the stabilizing role
of the Dark Matter (DM) halo throughout the galaxy, LSBGs are more stable to
many internal and external perturbations. Thus, only a weak spiral wave is
typically generated in a LSBG. This leads to some localized non-central
low-level SF, and, hence, to no appreciable changes of galaxy properties.
In contrast, in a HSB disk galaxy the strong perturbation of gas orbits leads
to global instability, resulting in the disturbed gas sinking into the
center of gravitational well (where it should eventually collapse and cause a
burst of SF).

The situation for this particular pair of gas-rich dwarfs is quite
reminiscent of these simulations.
Since the LSBD is significantly more massive than the BCG, the mutual effect
of their encounter will be enhanced for the BCG, and weaker for the LSBD.
However, currently available data on the properties of these galaxies are
too limited to make a more detailed comparison with galaxy interaction models.
\ion{H}{i} and H$\alpha$ mapping will help to determine some of their
parameters
(e.g., the full dynamic mass, gas density and SFR distribution),
and thus allow realistic modelling of this pair.

An additional argument for the effect of mutual interactions on
the enhanced SF in both galaxies is the fact of significant SF events in
each galaxy occurred on approximately the same timescale, which is comparable
to the characteristic times of the two galaxies.
All of them are on the order of a few hundred Myr. For the LSBD, we have
already discussed the ages of `old' component in composite models of its
$BVR$ colours. For HS 0822+3542 the age of the major SF episode is implied by
the detected Balmer absorptions in the underlying stellar disk (Izotov et al.,
in preparation). The time elapsed since the pericenter passage and strongest
interaction between the galaxies is approximately $\sim$200--300 Myr.
The development of
instabilities in a disturbed galaxy is delayed by a time comparable to
the revolution period. The latter is on the order of 100 Myr.
So, to a first approximation, the time since the strongest
interaction, the delay time for gas collapse, and the ages of the stellar
populations in the underlying disks of the LSBD and the BCG all are
comparable.

As an interacting galaxy,  HS~0822+3542 is similar to several other
XMD BCGs. Clear  evidence of tidal disturbance
was found for the optical counterpart of the large \ion{H}{i} cloud
Dw~1225+0152 (Salzer et al. \cite{Salzer91}, Chengalur et al.
\cite{Chengalur95}). The huge \ion{H}{i} cloud, hosting the pair of XMD
BCGs SBS~0335$-$052 E and W, also looks quite disturbed (Pustilnik et al.
\cite{PBTLI}). The synchronized SF episodes in the both BCGs are probably
the result of a tidal trigger. High-sensitivity \ion{H}{i} mapping
of the vicinities of other such galaxies would help to better understand the
trigger mechanisms of XMD BCG starbursts.

\section{Summary and conclusions}
\label{Conc}

In this work we present the physical parameters of a new LSB/dIrr galaxy
discovered in
the vicinity of one of the most metal-deficient BCGs, HS~0822+3542. Its small
relative velocity and projected distance
are consistent with a
scenario of tidal disturbance exerted on the BCG progenitor, which triggered
gas collapse and a subsequent SF burst.

The total $(B-V)$ and $(V-R)$ colours of SAO 0822+3545  are unusually blue
in comparison to the most of the known LSBD/dIrr galaxies. This
suggests that there was some recent elevated SF in this galaxy.
The most straightforward interpretation of the LSBD colours and $EW$ of
H$\alpha$ emission  imply that a substantial fraction of its stars are
young, with ages from $\sim$10 to $\sim$1~Gyr, depending on the model IMF
used.
Recent deep spectroscopy of HS~0822+3542 (Izotov et al., in preparation)
indicates the existence of a stellar population
with the ages of $50-100$ Myr. This is comparable to the age of
the suggested ``intermediate'' or ``old'' stellar populations in the LSBD.
Thus, in both components of the pair a substantial
fraction of stars formed recently, on a time scale of $\sim$100--200 Myr.
This suggests that this recent SF was triggered by the mutual tidal
action of these dwarfs.

This conclusion on the interaction-induced SF episodes in these dwarfs
is a preliminary one, based mainly on some general estimates and the
understanding that such an unusual pair of galaxies is unlikely to occur by
by chance.
Another important property of this pair is its high degree of isolation
from known bright galaxies. This presumably led to retarded star
formation and/or slow chemical evolution in HS~0832+3542, and a very low SFR
in SAO 0822+3545, as evidenced by their very high gas-mass fractions.

To better understand the evolutionary status of SAO 0822+3545, one needs
higher precision in both integrated colours and those of the
outermost parts of the LSBD. This can provide  direct confirmation of the
composite population model. H$\alpha$ images will help to determine the
spatial distribution of the recent SF episode, while colour-magnitude
diagrams for resolved stars will also provide direct age estimates for older
stellar populations.

From the results and discussion above we draw the following conclusions:

\begin{enumerate}
\item{BCG HS~0822+3542 has a physically associated LSB/dIrr galaxy at
 a projected distance of $\sim$11.4 kpc with a close radial velocity
  ($\Delta V \lesssim$25~\kms).}
\item{This LSB/dIrr galaxy, named SAO~0822+3545, has a low luminosity
 (M$_{\rm B}^{0} =$--12.85) and small optical size ($D_{25}\sim$1.5 kpc).
  Its  unusually blue integrated colours $(B-V)_\mathrm{0}$=0.08 and
  $(V-R)_\mathrm{0}$=0.14, for the case of a standard Salpeter IMF with
  $M_{\rm up}$=120$M$\sunn,
  indicate recent ($t\sim10-200$~Myr) SF episodes, in which stars comprising
  from $\sim$(6--7)\% to $\sim$100\% of the total stellar mass, have
  formed. The total stellar mass comprises $\sim$(7--27)\% of the galaxy's
  baryon  mass, depending on the age of the ``old'' stellar component.}
\item{Alternatively, if its IMF has $M_{\rm up}$ = 30~$M$\sunn\ and the
  same slope, the galaxy's properties can be explained by one component
  continuous SF during the last $\sim$1 Gyr.
  This, however, is less probable, considering the expected effects of
  a recent encounter.  }
\item{New,  high-S/N-ratio spectra of HS~0822+3542 yield an
  oxygen abundance of 12+$\log$(O/H)=7.44, consistent within the
  observational uncertainties with earlier measurements.
  The precision  of N/O is improved,  and the abundance of argon is
  derived for the first time.}
\item{In HS 0822+3542, for the first time the kinematic evidence has been
  found of a large ionized-gas  supershell, with a diameter of $\sim$0.5 kpc.
  The physical parameters of the supershell imply that the current SF episode
   in this BCG has lasted at least 5--9 Myr.}
\item{The dwarf pair is situated deep within the nearby Lynx-Cancer void, and
  thus has not been tidally disturbed by massive galaxies since its formation
  epoch.  All available  data are consistent with the hypothesis that
  we are witnessing the results of an interaction a few hundred Myr ago
  between two very gas-rich dwarf galaxies with significantly
  different masses and density distributions.}
\end{enumerate}

\begin{acknowledgements}

The authors thank J.Chengalur for permission to cite the GMRT results
from a paper in preparation.
We are grateful to Y.Izotov and N.Bergvall, who have critically read the
early versions of the paper and made many useful suggestions.
We are thankful to the anonymous referee, whose
criticism and useful suggestions helped to improve significantly the original
version of the paper.
We acknowledge partial support from INTAS grant 96-0500 and from the Russian
state program "Astronomy".
This research has made use of the NASA/IPAC Extragalactic
Database (NED), which is operated by the Jet Propulsion Laboratory,
California Institute of Technology, under contract with the National
Aeronautics and Space Administration, and  LEDA, the Lyon-Meudon
extragalactic database. The use of the Digitized Palomar Observatory Sky
Survey (DPOSS-II) and APM Database is gratefully acknowledged.

\end{acknowledgements}

\end{document}